\documentclass[aps,pre,reprint,showpacs,superscriptaddress]{revtex4-1}
\usepackage{amsmath,amssymb,physics}
\usepackage{graphicx}
\usepackage{dsfont}
\usepackage{bm}
\usepackage{xcolor}

\usepackage{epstopdf}
\usepackage{hyperref}

\newcommand{\kB}{k_{\rm\scriptscriptstyle B}}
\newcommand{\rms}{\rm\scriptscriptstyle}
\newcommand{\rhoL}{\rho_{\rm\scriptscriptstyle L}}
\newcommand{\rhoR}{\rho_{\rm\scriptscriptstyle R}}
\newcommand{\rhoB}{\rho_{\rm\scriptscriptstyle b}}

\DeclareMathOperator*{\argmin}{argmin}
\DeclareMathOperator*{\argmax}{argmax}
\newcommand{\vw}{v_{\rm w}}
\newcommand{\jst}{j_{\rm st}}
\newcommand{\varrhost}{\varrho_{\rm st}}

\newfont{\fortitle}{cmssbx11 scaled 1050}

\begin{document} 

\title[]{\fortitle Single-file transport in periodic potentials: The Brownian asymmetric simple exclusion process}

\author{Dominik Lips}
\email[]{dlips@uos.de}
\affiliation{Universit{\" a}t Osnabr{\" u}ck, Fachbereich Physik,
  Barbarastra{\ss}e 7, D-49076 Osnabr{\" u}ck, Germany}

\author{Artem Ryabov}
\email[]{rjabov.a@gmail.com}
\affiliation{Charles University, Faculty of Mathematics and Physics,
  Department of Macromolecular Physics, V Hole\v{s}ovi\v{c}k\'ach 2,
  CZ-18000 Praha 8, Czech Republic}
\affiliation{Centro de F{\' i}sica Te{\' o}rica e Computacional, Departamento de F{\' i}sica, 
Faculdade de Ci{\^ e}ncias, Universidade de Lisboa, Campo Grande P-1749-016 Lisboa, Portugal}

\author{Philipp Maass}
\email[]{maass@uos.de}
\affiliation{Universit{\" a}t Osnabr{\" u}ck, Fachbereich Physik,
  Barbarastra{\ss}e 7, D-49076 Osnabr{\" u}ck, Germany}

\date{18 June 2019; revised 18 September 2019} 

\begin{abstract}
Single-file Brownian motion in periodic structures is an important
process in nature and technology, which becomes increasingly amenable
for experimental investigation under controlled conditions. To explore
and understand generic features of this motion, the Brownian
asymmetric simple exclusion process (BASEP) was recently introduced.
This BASEP refers to diffusion models, where hard spheres are driven
by a constant drag force through a periodic potential.  Here, we
derive general properties of the rich collective dynamics in the
BASEP. Average currents in the steady state change dramatically with
the particle size and density. For an open system coupled to particle
reservoirs, extremal current principles predict various nonequilibrium
phases, which we verify by Brownian dynamics simulations.  For general
pair interactions we discuss connections to single-file transport by
traveling-wave potentials and prove the impossibility of current
reversals in systems driven by a constant drag and by traveling waves.

\end{abstract}

\maketitle  

\section{Introduction}
\label{sec:introduction}
Single-file transport refers to collective motion in strongly confined
geometries, where the particles cannot overtake each other.  In most
chemical and biophysical settings, single-file dynamics is severely
affected by thermal fluctuations and takes place far from
thermodynamic equilibrium.  In cell biology, examples of single-file
transport are the directed motion of motor proteins along microtubules
or actin filaments \cite{Lipowsky/etal:2001, Frey/Kroy:2005,
  Hille:2001}, ion migration through membrane channels
\cite{Hille:2001}, and protein synthesis by ribosomes
\cite{MacDonald/etal:1968}. With the steadily increasing quality of
experimental techniques capable to control and detect particle motion
on molecular scales, single-file transport in periodic (free) energy
landscapes becomes of increasing importance also for applications, as,
e.g., transport in carbon nanotubes \cite{Zeng/etal:2018}, zeolites
\cite{VanDeVoorde/Sels:2017}, mesoporous materials
\cite{Hartmann:2005, Humphrey/etal:2001}, and nanofluidic devices
\cite{Ma/etal:2015}.
 
A much investigated and well-understood feature in single-file
diffusion is the anomalous subdiffusion of a tracer particle in the
long-time limit \cite{Taloni/etal:2017}. This was first proven in the
mathematical literature \cite{Harris:1965} and later detected
experimentally in zeolites \cite{Hahn/etal:1996, Chmelik/etal:2018}
and in nanotubes \cite{Cheng/Bowers:2007, Dvoyashkin/etal:2014} by
using nuclear magnetic resonance techniques. Further direct observation was possible in
colloidal experiments by optical imaging \cite{Wei/etal:2000}.  The
aspect of subdiffusion was further elaborated in connection with
general theories of anomalous transport \cite{Taloni/etal:2010},
including descriptions in terms of fractional Brownian motion
\cite{Sanders/Ambjornsson:2012}, effects of initial and boundary
conditions \cite{Leibovich/Barkai:2013, Lizana/Ambjornsson:2008},
external force fields \cite{Barkai/Silbey:2009}, time-varying
potentials \cite{Ryabov/Chvosta:2011}, first-passage time
distributions \cite{Ryabov/Chvosta:2012, Ryabov:2013,
  Locatelli/etal:2016}, statistics of residence times
\cite{Krapivsky/etal:2015}, partial overtaking of particles
\cite{Mon/Percus:2002, Ooshida/etal:2018, Ahmadi/etal:2019}, and
large deviation functions for the position of a tracer
\cite{Krapivsky/etal:2014}.
 
As for collective transport properties, single-file motion has been
mainly investigated in lattice models, which reflect a periodic
structure in a coarse-grained manner.  These lattice models can be
considered as variants of the so-called asymmetric simple exclusion
process (ASEP) and have found applications in particular in the
modeling of biological traffic \cite{Schadschneider/etal:2010,
  Chou/etal:2011, Kolomeisky:2013, Appert-Rolland/etal:2015}.
 
The minimal ASEP model, where particles perform nearest-neighbor hops
between lattice sites with a bias in one direction and under the
constraint that only one particle can occupy a lattice site has become
a reference model for studying fundamental questions of statistical
physics out of equilibrium \cite{Derrida:1998, Schuetz:2001}.  For
this model, exact results for microstate distributions in
nonequilibrium steady states could be obtained
\cite{Blythe/Evans:2007}. When coupled to particle reservoirs, three
different phases of nonequilibrium steady states appear in dependence
of the reservoir densities \cite{Krug:1991,
  Parmeggiani/etal:2003}. Studies with nearest-neighbor interactions
between particles showed richer phase diagrams
\cite{Antal/Schuetz:2000, Dierl/etal:2011}. They led to a
clarification of the role of system-reservoir couplings in open
systems for the topology of nonequilibrium phase diagrams and of the
meaning of particle-hole symmetry \cite{Dierl/etal:2012,
  Dierl/etal:2013}. This clarification turned out to be useful also
for understanding collective particle dynamics in lattice models with
time-varying site energies \cite{Dierl/etal:2014}.
 
Many further interesting results were reported for the ASEP, as, e.g.,
singularities in large deviation functions for time-averaged currents
\cite{Bertini/etal:2005, Lazarescu:2015, Baek/etal:2017}, and
condensation transitions for nonuniform hopping rates
\cite{Evans:1996, Concannon/Blythe:2014}.  More recently, new
universality classes in the hydrodynamic limit of nonlinear
hydrodynamics were discovered for multi-lane variants of the model
\cite{Popkov/etal:2015}. Predictions for long-time tails could be
proven for a specific microscopic model \cite{Chen/etal:2018}.
 
As the hopping transitions in the ASEP can be considered to reflect
rare events of thermally activated barrier crossings, one may
conjecture that continuous single-file motion in a periodic potential
exhibits features similar to the ASEP. However, in a recent study
\cite{Lips/etal:2018} we showed that a much richer behavior occurs in
continuous Brownian motion because of additional length scales
associated with particle-particle interactions. The
  simplest class of models is that of hardcore interacting particles,
which we refer to as the Brownian asymmetric simple exclusion
process (BASEP). In this BASEP, hard rods with diameter $\sigma$ are
driven through a periodic potential with wavelength $\lambda$ by a
constant drag force $f$.
 
The BASEP is particularly interesting in connection with recent
experiments utilizing advanced techniques of microfluidics and optical
and/or magnetic micromanipulation \cite{Arzola/etal:2017,
  Skaug/etal:2018, Schwemmer/etal:2018, Stoop/etal:2018,
  Misiunas/Keyser:2019}, which includes setups where the particles are
not driven by a constant drag force but a traveling-wave potential
\cite{Straube/Tierno:2013}. This is because the Brownian motion of a
particle in a traveling-wave potential $U(x-\vw t )$ is mapped onto
that in a periodic potential $U(x)$ with a constant drag force $f
=-\vw/\mu$ after a coordinate transformation $x \to x-\vw t$; $\mu$ is
the bare mobility of the particles.
 
Furthermore, the BASEP should allow one to understand under which
conditions a coarse-grained description in terms of lattice models
will be appropriate.  Generally, a more detailed understanding of the
connection between the BASEP and corresponding lattice models is
necessary to explain why certain effects are seen in one description
but not in the other. For example, steady-state currents
  opposite to the external bias, so-called current-reversals, were
reported for lattice models \cite{Jain/etal:2007, Slanina:2009JSP,
  Slanina:2009PRE, Chaudhuri/Dhar:2011, Dierl/etal:2014} and recently
seen experimentally in a rocking Brownian motor
\cite{Schwemmer/etal:2018}, but they were not found in an analogous
setting with continuous-space dynamics \cite{Chaudhuri/etal:2015}.
Current reversals in space-continuous models were reported earlier for
a constant and ``flashing'' asymmetric sawtooth-shaped external
potential \cite{Derenyi/Viscek:1995, Derenyi/Ajdari:1996}, and in a
recent work with time-discontinuous driving of a single potential
barrier along a ring \cite{Rana/etal:2018}.  In the BASEP, a barrier
reduction and an exchange symmetry effect were identified as decisive
mechanisms for the characteristics of single-file transport, but it is
yet unclear to which extent these effects have a counterpart in
lattice descriptions. In this study, we will provide further insight
into these issues.
 
A further goal of this study is to gain an extended description of the
average steady-state current $\jst$ in the BASEP and an explicit
verification of the phases of nonequilibrium steady states in an open
system coupled to particle reservoirs. We believe that this provides a
useful basis for future investigations of model variants where further
details can be included to capture specific experimental conditions.
 
As for the extended description of the average steady-state current
$\jst$ (which we often refer to as ``the current'' in the following),
we note that our results reported in Ref.~\cite{Lips/etal:2018} focused on
a limited regime of particle densities $\rho$, which we defined as the
(dimensionless) filling factor of the potential wells. The maximal
possible $\rho$ is equal to $\lambda/\sigma$ and the yet unexplored
regime is that of high densities in the range $1 <
\rho\le\lambda/\sigma$. In fact, by making use of a mapping of
currents for different particle sizes, we give a description of
$\jst=\jst(\rho,\sigma)$ for all particle sizes and densities.

Moreover, we investigate the influence of the temperature on $\jst$,
and prove that current reversals cannot occur in closed systems driven
by a constant drag force or a traveling wave.  This leads to rigorous
upper bounds of $|\jst|$ and of $|\overline{\jst^{\rms TW}}|$, where
$\overline{\jst^{\rms TW}}$ denotes the period-averaged steady-state
current in a traveling-wave driven system.
 
Our simulations are supported by an analytical treatment extending the
derivations in the supplemental material of
Ref.~\cite{Lips/etal:2018}. This yields an approximate expression for
the current in the linear-response regime, which shows qualitative
agreement with the simulation results. We show that quantitative
deviations are mainly due to the neglect of an interaction-mediated
effective drift term, which stems from the interplay between the
external periodic potential and interparticle interactions in the
steady state. This mean interaction force has a clear physical
interpretation, yet its quantitative analytical description is
challenging.

In the open BASEP coupled to particle reservoirs, we introduce a
simple scheme of particle injection and ejection, which allows us to
demonstrate four of the five possible phases predicted by the extremal
current principles in Ref.~\cite{Lips/etal:2018}. The missing phase
corresponds to a small region in the phase diagram appearing at very
high reservoir densities. All simulated phases are demonstrated by
corresponding density profiles.

As our results for analytical expressions of the current, current
reversals and current bounds are valid for quite general pair
interactions, we first discuss collective Brownian single-file
transport for this wider class of systems.  From
Sec.~\ref{subsec:recurrent-dynamics} onward, we focus on the BASEP with
hardcore-interacting particles.

\section{Single-file Brownian motion}
\label{sec:sfbm}
We consider single-file Brownian motion of $N$ particles in a periodic
potential $U(x)=U(x+\lambda)$.  The particles are driven by a constant
drag force $f$ and interact via pair forces $f^{(2)}(x,y)$,
i.e., $f_i^{\rm \scriptscriptstyle int}=\sum_{j\ne i}f^{(2)}(x_i,x_j)$
is the interaction force on the $i$th particle. For overdamped
Brownian motion, the particle dynamics are described by the Langevin
equations
\begin{align}
\frac{\dd x_i}{\dd t} = 
\mu \left(f+f_i^{\rm \scriptscriptstyle int}-
\frac{\partial U(x_i)}{\partial x_i}\right) + \sqrt{2D} \, \eta_i(t) \,,
\label{eq:langevin}
\end{align}
where $\mu$ and $D=\mu \kB T$ are the bare mobility and diffusion
coefficient, and $\kB T$ is the thermal energy. The $\eta_i(t)$ are
independent and $\delta$-correlated Gaussian white noise processes
with zero mean and unit variance, $\langle \eta_i \rangle =0$ and
$\langle \eta_i(t) \eta_j(t') \rangle = \delta_{ij} \delta(t-t')$.

\begin{figure}[t!]
\includegraphics[width=\columnwidth]{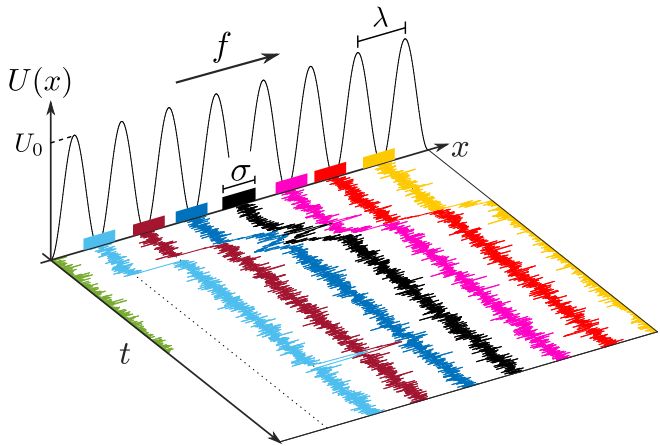}
\caption{Illustration of the Brownian asymmetric simple exclusion
  process, where hardcore interacting particles of size $\sigma$ are
  driven through a $\lambda$-periodic potential $U(x)$ with amplitude
  $U_0\gg \kB T$ by a constant drag force $f$. The example of
  eight stochastic trajectories of neighboring particles was obtained from 
  Brownian dynamics simulations 
  in a system with $\sigma=0.5\lambda$ at a density $\rho=0.8$.}
\label{fig:model}
\end{figure}

For the BASEP, the hardcore interactions imply the boundary conditions
$|x_{i} - x_{j}| \geq \sigma$, i.e., overlaps between neighboring
particles are forbidden. Taking into account these boundary conditions,
the interaction force $f_i^{\rm \scriptscriptstyle int}$ can be set to
zero in Eq.~\eqref{eq:langevin}.  We define the density as a
(dimensionless) filling factor of the potential wells, i.e., by
$\rho=N/M$, where $M$ denotes the total number of periods of $U(x)$.
The system length is $L=M\lambda$ and the number density is
$\rho/\lambda$ with the upper bound $1/\sigma$.  The model is sketched
in Fig.~\ref{fig:model} for the cosine periodic potential $U(x)$
discussed thoroughly in Secs.~\ref{sec:current_specific_example} and
\ref{sec:open-basep}.

In the following we will distinguish between the closed BASEP, where
periodic boundary conditions are applied with particle coordinates in
sequential order $x_{1} \leq x_{2}\leq\ldots \leq x_N$ \footnote{To
  implement the periodic boundary conditions, we assume an ordered
  initial configuration $0 \leq x_{1} \leq x_{2} \ldots \leq x_N < L$,
  and introduce two fictive particles with enslaved coordinates
  $x_0=x_N-L$ and $x_{N+1} = x_1+L$, which implies $x_N - x_1 <
  L-\sigma$. This implementation, where the $x_i$ can assume any real
  value, is convenient for discussing transformation properties of the
  Langevin equations. Alternatively, one could consider the particle
  positions to be confined to a ring of size $L$, corresponding to a
  mapping $x_i \to x_i \, \mathrm{mod} \, L$. However, then the
  sequential order of the particles can no longer be expressed by
  $x_{1} \leq x_{2} \ldots \leq x_N$ in the course of time.}, and the
open BASEP, where the left and right boundaries are coupled to
particle reservoirs with in general different densities. In the closed
BASEP, the drag force $f$ leads to a steady state with a constant
particle current $\jst$ and a $\lambda$-periodic local density profile
$\varrhost(x)$.  The dependence of the steady-state current $\jst$ on
$\rho$ and $\sigma$ is denoted explicitly, i.e., $\jst=\jst(\rho,
\sigma)$, while other dependencies on $T$, $f$ etc.\ are omitted in
the notation. In the open BASEP, specification of the way of how
particles are exchanged with the reservoirs is a subtle issue, which
will be discussed in Secs.~\ref{subsec:phasetrans} and
\ref{subsec:simulated_phases}.

In simulations of the BASEP, we used numerical algorithms specifically
developed for Brownian dynamics of hard-sphere systems
\cite{Tao/etal:2006, Scala/etal:2007,Scala:2012,
  Behringer/Eichhorn:2012}.  These algorithms use the Euler method and
differ in the implementation of the hardcore (excluded volume)
constraints. Specifically, we applied the two algorithms developed in
Refs.~\cite{Scala:2012} and \cite{Behringer/Eichhorn:2012}. Our
results are not affected by the choice of any of these algorithms, and
they showed agreement with exact analytical findings for specific
cases.

\section{Current in closed systems: General results}
\label{sec:closed-systems}

\subsection{Continuity equation and steady-state current}
\label{subsec:jst}
The conservation of the particle number in a closed
  system is described by the continuity equation
\begin{equation}
\frac{\partial \varrho (x, t)}{\partial t} = -\frac{\partial j(x, t)}{\partial x} \, ,
\label{eq:one-particle-equation}
\end{equation}
where $\varrho(x,t)=\langle\sum_{i=1}^N \delta(x-x_i)\rangle$ is the
local density and $\langle \ldots \rangle$ refers to the ensemble
average at time $t$ (for an arbitrary given initial condition).  The
local particle current is \cite{Lips/etal:2018}
\begin{equation}
j(x,t) = \mu [f^{\rm \scriptscriptstyle ext}(x)+ f^{\rm
    \scriptscriptstyle int}(x,t)]\varrho(x,t)
- D \frac{\partial \varrho(x,t)}{\partial x} \, ,
\label{eq:particle-current-definition}
\end{equation}
where $f^{\rms ext}(x)=f-\partial U(x)/\partial x$, and
$f^{\rms int}(x,t)$ is the mean force at position $x$ at time
$t$ due to the interactions. With the two-particle density
$\varrho^{(2)}(x, y, t)=\langle \sum_{i=1}^N \sum_{k=1, k \neq i}^N
\delta(x-x_i)\delta(y-x_k) \rangle$, this force can be written as
\begin{align}
f^{\rms int}(x,t) 
&= \frac{1}{\varrho(x,t)} \int_0^L \dd y\, f^{(2)}(x,y)\varrho^{(2)}(x, y, t) \nonumber \\ 
&= \int_0^L \dd y\, f^{(2)}(x,y) \varrho(y|x; t) \, ,
\label{eq:mean-interaction-force}
\end{align}
where $\varrho(y|x; t)=\varrho^{(2)}(x, y, t)/\varrho(x,t)$ is the
local density at position $y$ at time $t$ under the condition that a
particle is at position $x$ at time $t$.

For hardcore interactions, an exact treatment based on
the many-particle Smoluchowski equation given in the supplemental
material of Ref.~\cite{Lips/etal:2018} leads to
\begin{equation}
f^{\rm \scriptscriptstyle int}(x,t) = \kB T\,
\frac{\varrho^{(2)}(x, x-\sigma, t) - \varrho^{(2)}(x, x+\sigma, t)}{\varrho(x,t)} \, .
\label{eq:def-mean-interaction-force}
\end{equation}
That expression follows also when formally setting $f^{(2)}(x,y)=\kB T
[\delta(y-(x-\sigma)) - \delta(y-(x+\sigma))]$ in
Eq.~\eqref{eq:mean-interaction-force}.  Intuitively, this can be
understood as follows: For a particle at position $x$ there is a
positive and a negative force on contacts with other particles at
positions $x-\sigma$ and $x+\sigma$ that correspond to the two
$\delta$-functions. The amplitude in front of the $\delta$-functions
must be an energy on dimensional reasons, for which $\kB T$ is the
only relevant scale.

In a steady state, the density profile $\varrho(x,t)$ is
time-independent, $\varrho(x,t)=\varrhost(x)$, and the current
homogeneous, $j(x,t)=\jst$. For periodic boundary conditions, the
$\lambda$ periodicity of the external force $f^{\rms ext}$ entails
that the steady-state solution is as well $\lambda$ periodic,
i.e., $\varrhost(x)=\varrhost(x+\lambda)$. Dividing
Eq.~(\ref{eq:particle-current-definition}) by $\varrhost(x)$ and
integrating over one period, we obtain for the steady-state current
\begin{equation}
\jst(\rho, \sigma)=
\frac{\mu\left(f+\frac{1}{\lambda} \int_0^\lambda\dd x f^{\rms int}_{\rm st}(x)\right)}
{\displaystyle \frac{1}{\lambda} \int_0^\lambda\dd x \frac{1}{\varrhost(x)}} 
=\frac{\mu\left(f + \overline{f^{\rms int}_{\rm st}}\right)}{\overline{\varrhost^{-1}}}\,,
\label{eq:ness-current}
\end{equation}
where
$\overline{\ldots\phantom|\hspace{-1.5ex}}=\lambda^{-1}\int_0^\lambda
\dd x\ldots$ denotes a period-averaging.

\subsection{Relation to traveling-wave driving}
\label{subsec:TW}
Overdamped Brownian motion in a traveling-wave potential with wave
velocity $v_{\rm w}$ is described by the Langevin equations
\begin{equation}
\frac{\dd x_i}{\dd t} = -\mu
\frac{\partial U(x_i - v_{\rm w} t)}{\partial x_i} 
+\mu f_i^{\rm \scriptscriptstyle int}+ \sqrt{2D} \, \eta_i(t)\,.
\label{eq:langevin-TW}
\end{equation}
If the pair interaction force $f^{(2)}(x,y)$ is a function of the pair
distance $(x-y)$ only, then the Galilean transformation $x_i\to x'_i = x_i
- v_{\rm w} t$ reduces these equations to Eqs.~\eqref{eq:langevin}
with the constant drag force $f=-v_{\rm w}/\mu$ (the primed
coordinates corresponding to the system with constant drag force).
Note that also hardcore constraints remain invariant under the
Galilean transformation, $|x_{i+1}-x_i|= |x'_{i+1}-x'_{i}|\geq\sigma$.

The correspondence implies the relations 
\begin{subequations}
\label{eq:rho-j-transform}
\begin{align}
\varrho^{\rms TW}(x,t)&=\varrho(x-\vw t,t)\label{eq:rho-j-transform-a}\,,\\
j^{\rms TW}(x,t)
&=j(x-\vw t,t)+\vw\varrho(x-\vw t,t)\label{eq:rho-j-transform-b}
\end{align}
\end{subequations}
between local particle densities and currents, where the superscript
``TW'' marks the quantities in the traveling-wave system.
Equation~\eqref{eq:rho-j-transform-a} is quite obvious and it follows
formally by transforming to the coordinates $x_i'(t)=x_i(t)-\vw t$ of
the comoving frame,
\begin{align}
\varrho^{\rms TW}(x,t)&=\left\langle\sum_{i=1}^N \delta(x-x_i(t))\right\rangle\nonumber\\
&\hspace{-2em}=\left\langle\sum_{i=1}^N \delta(x-\vw t-x_i'(t))\right\rangle=\varrho(x-\vw t,t)\,.
\label{eq:rho-transform}
\end{align}
In the same manner, one finds (and corresponding relations hold for
many-particle densities of higher order)
\begin{align}
\varrho^{(2),{\rms TW}}(x,y,t)&=\varrho^{(2)}(x-\vw t,y-\vw t,t)\,.
\label{eq:rho2-transform}
\end{align}
Given Eqs.~\eqref{eq:rho-transform} and \eqref{eq:rho2-transform}, the
local mean interaction forces in the two systems satisfy an analogous
relation [cf.\ Eq.~\eqref{eq:mean-interaction-force}],
\begin{align}
f^{\rm{int},\rms TW}(x,t)=f^{\rm int}(x-\vw t,t)\,.
\end{align}
For the transformation between the local currents we thus obtain
[cf.\ Eq.~\eqref{eq:particle-current-definition}]
\begin{align}
j^{\rms TW}(x,t)&=
\mu\left(f^{\rm{int},\rms TW}(x,t)
-\frac{\partial U(x-\vw t)}{\partial x}\right)\varrho^{\rms TW}(x,t)\nonumber\\
&\phantom{=}-D\frac{\partial \varrho^{\rms TW}(x,t)}{\partial x}\nonumber\\
&\hspace{-3em}=\mu\left(f^{\rm int}(x-\vw t,t)-
\frac{\partial U(x-\vw t)}{\partial x}\right)\varrho(x-\vw t,t)\nonumber\\
&\phantom{=}-D\frac{\partial \varrho(x-\vw t,t)}{\partial x}\nonumber\\
&=j(x-\vw t,t)-\mu f \varrho(x-\vw t,t)\nonumber\\
&=j(x-\vw t,t)+\vw \varrho(x-\vw t,t)\,,
\label{eq:jtw}
\end{align}
which agrees with Eq.~\eqref{eq:rho-j-transform-b}.  For steady
states, Eqs.~\eqref{eq:rho-j-transform-a} and
\eqref{eq:rho-j-transform-b} become
\begin{subequations}
\label{eq:rho-j-transform-steady}
\begin{align}
\varrhost^{\rms TW}(x,t)&=\varrhost(x-\vw t)\label{eq:rho-j-transform-steady-a}\,,\\
\jst^{\rms TW}(x,t)
&=\jst+\vw\varrhost(x-\vw t)\,.\label{eq:rho-j-transform-steady-b}
\end{align}
\end{subequations}

The relations given in this section can be of particular interest in
experimental setups with optical or magnetic tweezers, where
traveling-wave potentials represent a feasible way to mimic the
constant drift force \cite{Straube/Tierno:2013}. They are also of
interest in the following discussion of current reversal.

\subsection{Absence of current reversals and implications for current bounds}
\label{subsec:current-bounds}
In systems with a constant drag force $f$, a current reversal is not
possible in the steady state, i.e., $\jst$ must have the same sign as
$f$. This can be expected from the second law of thermodynamics, which
forbids the system to continuously extract heat from its environment
and to perform work against the external force. In systems driven by a
traveling wave, the impossibility of current reversals is less
expected, in particular because such reversals were seen in
corresponding lattice models \cite{Chaudhuri/Dhar:2011,
  Dierl/etal:2014}.  An absence of current reversals in traveling-wave
systems with continuous space dynamics was conjectured in
Ref.~\cite{Chaudhuri/etal:2015} based on Brownian dynamics simulations
for different pair potentials and a perturbative expansion of the
single-particle density around its period-averaged value. Here we give
rigorous proofs for the BASEP for both types of driving.

To this end we consider the entropy production
\begin{align}
\frac{\dot{S}(t)}{\kB}&=-
\frac{\partial}{\partial t} \int_\Omega \dd^N\!x\, p_{\scriptscriptstyle N} \ln p_{\scriptscriptstyle N}
=-\int_\Omega \dd^N\!x\,\frac{\partial p_{\scriptscriptstyle N} }{\partial t}\ln p_{\scriptscriptstyle N}
\label{eq:S1}
\end{align}
in the system, where $p_N=p_N(x_1,\ldots,x_N,t)$ is the $N$-particle
joint probability density and $\Omega$ specifies the allowed
configuration space (for hardcore interactions, this is restricted
due to the conditions $|x_{i+1}-x_i|\ge\sigma$).  Using the
Smoluchowski equation $\partial p_N/\partial t=-\sum_{i=1}^N \partial
J_i/\partial x_i$ with the probability currents
\begin{equation}
J_i=J_i(x_1, \ldots, x_{\scriptscriptstyle N}, t) 
=\mu f^{\rms ext}(x_i)p_{\scriptscriptstyle N}
-D \frac{\partial p_{\scriptscriptstyle N}}{\partial x_i} \, ,
\label{eq:ith-smoluchowski-current}
\end{equation}
the time derivative of $p_N$ in Eq.~\eqref{eq:S1} can be written as
\begin{align}
\frac{\dot{S}(t)}{\kB}&=\sum_{i=1}^N \int_\Omega \dd^N\!x\, \frac{\partial J_i}{\partial x_i} \ln p_{\scriptscriptstyle N}
=-\sum_{i=1}^N \int_\Omega \dd^N\!x\,  J_i \frac{\partial \ln p_{\scriptscriptstyle N}}{\partial x_i} 
\nonumber\\
&= \sum_{i=1}^N \int_\Omega \dd^N\!x\, \frac{J_i^2}{Dp_{\scriptscriptstyle N}}
-\frac{1}{\kB T}\sum_{i=1}^N\int_\Omega \dd^N\!x\,  J_i f^{\rms ext}(x_i)\nonumber\\
&=\frac{\dot S_{\rm tot}(t)}{\kB}-\frac{\dot S_{\rm m}(t)}{\kB}\,.
\label{eq:S2}
\end{align}
Here we took into account the periodic boundary conditions in the
partial integration and expressed $\partial\ln p_{\scriptscriptstyle
  N}/\partial x_i$ using Eq.~\eqref{eq:ith-smoluchowski-current}.  In
the last step, we identified the resulting two terms as the production
of total entropy and entropy in the surrounding medium in the
framework of stochastic thermodynamics \cite{Seifert:2012}.  Note that
the total entropy production is positive out of equilibrium
($\sum_{i=1}^N J_i^2>0$) in agreement with the second law of
thermodynamics.

In the nonequilibrium steady state, the entropy production in the
system is zero, $\dot S=0$, and the (time-independent) total entropy
production $\dot S_{\rm tot}>0$ equals the entropy production $\dot
S_{\rm m}$ in the medium.  Because the partial current $j_i(x,t)$ of
the $i$th particle is $J_i$ integrated over all but the $i$th
coordinate, i.e., $j_i(x,t)=\int_{\Omega}\dd^N\!x\,
J_i(x_1,\ldots,x_N,t)\delta(x-x_i)$, and $j_i(x,t)=\jst/N$ in the
steady state, we obtain from Eq.~\eqref{eq:S2}
\begin{align}
\dot S_{\rm tot}&=\dot S_{\rm m}=\frac{\jst}{T}\int_0^L\dd x\, f^{\rms ext}(x)
\nonumber\\
&=\frac{\jst M}{T}\int_0^\lambda\dd x\left(f-\frac{\partial U(x)}{\partial x}\right)=\frac{\jst fL}{T}>0\,.
\label{eq:Sm2}
\end{align}
Here we have used the periodicity of the potential and
$L=M\lambda$. The total entropy production has the Onsager form of
current $\jst$ times the thermodynamic force $fL/T$. Its positive
definiteness implies that $\jst$ and $f$ must have the same sign.

Likewise, it can be shown that the total entropy production in the
steady state of a traveling-wave driven system obeys an analogous
Onsager type relation, if period-averaged quantities are considered.
Using the general expression $\dot S_{\rm
  m}=T^{-1}\sum_{i=1}^N\int_\Omega \dd^N\!x\, J_i f^{\rms ext}(x_i)$
for the entropy production in the medium, see Eq.~\eqref{eq:S2}, we
obtain for a traveling-wave system
\begin{align}
\dot S_{\rm m}^{\rms TW}(t)&=
-\frac{1}{T}\sum_{i=1}^N\int_{\Omega}\dd^N x\, J_i^{\rms TW}\frac{\partial U(x_i-\vw t)}{\partial x_i}\\
&=-\frac{1}{T}\sum_{i=1}^N\int_0^L\dd x_i\, \frac{\partial U(x_i-\vw t)}{\partial x_i}\,j_i^{\rms TW}(x_i,t)\,.
\nonumber
\end{align}
Time averaging of the total entropy production $\dot S_{\rm tot}^{\rms
  TW}=\dot S_{\rm m}^{\rms TW}+\dot S^{\rms TW}$ over one period
$\tau=\lambda/\vw$ in the steady state yields, under consideration of
$S(t+\tau)=S(t)$ and $j_{{\rm st},i}^{\rms TW}(x,t)=j^{\rms TW}_{\rm
  st}{(x,t)}/N=\jst+\vw \varrho_{\rm st}(x-\vw t)$
[cf.\ Eq.~\eqref{eq:rho-j-transform-steady-b}],
\begin{align}
\overline{\dot S_{\rm tot}^{\rms TW}}&=
-\frac{M}{T\tau}\int_0^\tau\!\!\!\!\dd t\!\!\int_0^\lambda\!\!\!\dd x\,\frac{\partial U(x\!-\!\vw t)}{\partial x}\,
[\jst+\vw\varrho_{\rm st}(x\!-\!\vw t)]\nonumber\\
&=\frac{M\vw}{T}\int_0^\lambda\dd x\, \left(-\frac{\dd U(x)}{\dd x}\varrho_{\rm st}(x)\right)>0\,.
\label{eq:stot}
\end{align}
Here, $\overline{\ldots\phantom|\hspace{-1ex}}=\tau^{-1}\int_0^\tau
\dd t\ldots$ denotes the time averaging.  Using
Eq.~\eqref{eq:particle-current-definition} in the steady state with
$\mu f=-\vw$, the integrand in the last expression of
Eq.~\eqref{eq:stot} can be written as
\begin{align}
-\frac{\dd U(x)}{\dd x}\varrho_{\rm st}(x)&=
\frac{1}{\mu}\,[\jst+\vw\varrho_{\rm st}(x)]-f^{\rms int}_{\rm st}(x)\varrho_{\rm st}(x)\nonumber\\
&\phantom{=}+{\kB T}\,\frac{\dd\varrho_{\rm st}(x)}{\dd x}\,.
\end{align}
Inserting this into Eq.~\eqref{eq:stot} gives, when utilizing the
$\lambda$ periodicity of $\varrho_{\rm st}(x)$, and taking into
account that $\overline{\jst^{\rms TW}}=\jst+\vw\overline{\varrho_{\rm
    st}}$ [cf.\ Eq.~\eqref{eq:rho-j-transform-steady-b}]
\begin{align}
\overline{\dot S_{\rm tot}^{\rms TW}}
=\frac{\vw L}{\mu T}
\left[\overline{j_{\rm st}^{\rms TW}}-\mu\overline{f^{\rms int}_{\rm st}(x)\varrho_{\rm st}(x)}\right]>0\,.
\end{align}
Note that in the system with constant drag force, the bar always
refers to a spatial period averaging, while in the corresponding
traveling-wave system, time  and space averaging over one period
yield the same results in the steady state.

The average $\overline{f^{\rms int}_{\rm st}(x)\varrho_{\rm st}(x)}$
vanishes for pair interaction forces obeying the principle of actio
and reactio, $f^{(2)}(x,y)=-f^{(2)}(y,x)$ (for hardcore interactions
one can use $f^{(2)}(x,y)=\kB T[\delta(y-x+\sigma)-\delta(y-x-\sigma)]$, see the discussion after
Eq.~\eqref{eq:def-mean-interaction-force}):
\begin{align}
\overline{f^{\rms int}_{\rm st}(x)\varrho_{\rm st}(x)}
&=\frac{1}{\lambda}\int_0^\lambda\dd x\,\varrho_{\rm st}(x) f^{\rms int}_{\rm st}(x)\nonumber\\
&=\frac{1}{L}\int_0^L\dd x\,\varrho_{\rm st}(x) f^{\rms int}_{\rm st}(x)\nonumber\\
&\hspace{-3em}=\frac{1}{L}\int_0^L\dd x
\int_0^L\dd y\, \varrho_{\rm st}^{(2)}(x,y)f^{(2)}(x,y)=0\,.
\end{align}
Here we have inserted $f^{\rms int}_{\rm st}(x)$ from
Eq.~\eqref{eq:mean-interaction-force} and used that the double
integral is zero because the two-particle density is a symmetric
function, $\varrho_{\rm st}^{(2)}(x,y)=\varrho_{\rm
  st}^{(2)}(y,x)$. Hence,
\begin{align}
\overline{\dot S_{\rm tot}^{\rms TW}}
=\frac{\overline{\jst^{\rms TW}}\vw L}{\mu T}>0\,.
\label{eq:stot-tw}
\end{align}
This has an Onsager form fully analogous to
Eq.~\eqref{eq:Sm2} but now for the period-averaged quantities.  It
follows that $\overline{j_{\rm st}^{\rms TW}}$ and $\vw$ must have the
same sign for traveling-wave potentials. The above derivations can be 
generalized to systems with short-range conservative interaction forces.

The absence of current reversals imply upper bounds for the magnitudes
of the currents: From Eq.~\eqref{eq:stot-tw} we obtain
$\overline{\jst^{\rms TW}}\vw=-(\jst-\mu f\overline{\varrhost})\mu
f>0$ and with $\jst f>0$ this implies $|\jst|<\mu |f|
\overline{\varrhost}=\mu |f|\rho/\lambda$. From Eq.~\eqref{eq:Sm2} it
follows $\jst f=-(\overline{\jst^{\rms
    TW}}-\vw\overline{\varrhost})\vw/\mu>0$ and with
$\overline{\jst^{\rms TW}}\vw>0$ this implies $|\overline{\jst^{\rms
    TW}}|<\overline{\varrhost}|\vw|=\rho|\vw|/\lambda$.  These bounds
can be expected: In a system with constant drag force, it means that
the magnitude of the current can never exceed that of independent
particles in a flat potential ($U=0$). In a traveling-wave system, it
means that the magnitude of the current can never exceed that of
particles coherently co-moving with the wave.

\subsection{Recurrent dynamics in periodicity intervals of the particle diameter in the BASEP}
\label{subsec:recurrent-dynamics}
In the BASEP, the complete range of densities (filling factors) $\rho$
and particle diameters $\sigma$ covers the range $0 \leq \sigma <
\infty$ and $0 \leq \rho \leq \lambda/\sigma$. We now discuss that for
a closed system with periodic boundary conditions, it is sufficient to
know the behavior in the reduced range $0 \leq \sigma < \lambda$
because of a recurrent dynamics in successive intervals separated by
integer multiples of the diameter, i.e., for $m\lambda < \sigma \leq
(m+1)\lambda$, $m=1,2,\ldots$ \cite{Lips/etal:2018}.

The reason for this is as follows. Let us consider a system with
particle diameter $\sigma' > \lambda$ and denote by
$m=\mathrm{int}(\sigma'/\lambda)$ the integer number of periods
fitting into $\sigma'$. By applying the coordinate transformation
$x_i' \to x_i = x_i'-im\lambda$ to the Langevin equations
\eqref{eq:langevin}, the external forces remain the same, $f^{\rms
  ext}(x_i') = f^{\rms ext}(x_i+im\lambda) = f^{\rms ext}(x_i)$, and
the hardcore constraints become $|x_{i+1} - x_i| \geq \sigma'
-m\lambda = \sigma$. Moreover, the confinement $(x_N' - x_1') <
(L'-\sigma')$ transforms into $(x_N - x_1) < L' -(N-1)m\lambda -
\sigma' = (L' - Nm\lambda) - \sigma$, corresponding to a change $L'
\to L=L'-Nm\lambda$ of the system length.

Hence there is a one-to-one mapping of probabilities of paths
$\{x_i'(t)\}$ and $\{x_i(t) = x_i'(t) -im\lambda\}$ between a system
with $\sigma'$ and $L'$ and a system with $\sigma=\sigma' - m\lambda$
and $L=L'-Nm\lambda$ (for fixed particle number $N$),
\begin{equation}
P\big[\{x_i'(t)\} ; L', \sigma'\big] = P \big[\{x_i(t)\} ; L, \sigma \big] \, .
\label{eq:path-probabilites}
\end{equation}
The arguments after the semicolon indicate the respective system
length and particle diameter. Knowing the behavior of an observable
for diameters $0\le\sigma<\lambda$ and for all densities $0 \leq \rho
\leq \lambda/\sigma$, one can infer the behavior of related
observables for all particle diameters in the range
$m\lambda\le\sigma<(m+1)\lambda$ for $m\ne0$.

In the thermodynamic limit, relations connecting systems at different
$L=M\lambda$ can be rewritten in terms of relations connecting systems
at different $\rho$. The change in system size then transfers into a
change $\rho' = N/M' \to \rho =N/M=\rho'/(1-\rho' m)$ of the density.
Specifically, for the relevant steady-state quantities considered
below, namely the local density $\varrhost(x;\rho, \sigma)$ and the
current $\jst(\rho, \sigma)$, we obtain [with
  $m=\mathrm{int}(\sigma'/\lambda)$]
\begin{subequations}
\label{eq:particle-diameter-mapping}
\begin{align}
\varrhost(x;\rho', \sigma') &= \frac{\rho'}{\rho} \, \varrhost(x;
\rho, \sigma) 
\label{eq:particle-diameter-mapping-a} \\ 
&= (1-m\rho') \, \varrhost\left(x; \frac{\rho'}{1-m\rho'},\sigma'-m\lambda\right)\,, 
\nonumber\\
\jst(\rho',\sigma') &= (1-m\rho')\,\jst\left(\frac{\rho'}{1-m\rho'},\sigma'-m\lambda\right)\,.
\label{eq:particle-diameter-mapping-b}
\end{align}
\end{subequations}
Equation \eqref{eq:particle-diameter-mapping-b} can be derived as
follows: The Langevin equations \eqref{eq:langevin} remain unchanged
under the coordinate transformation $x_i' \to x_i = x_i'-im\lambda$,
which implies equal mean velocities $v_{\rm st}$ in the original and
transformed system. Then, using $\jst'/\rho' = v_{\rm st}=\jst/\rho$
and Eq.~\eqref{eq:particle-diameter-mapping-a} yields
Eq.~\eqref{eq:particle-diameter-mapping-b}.

Equations~\eqref{eq:particle-diameter-mapping} imply a
commensurability effect for particle diameters equal to integer
multiples of the periodicity length $\lambda$, where the dynamics
becomes reducible to that for $\sigma=0$. In this case of hardcore
interacting point particles, it is obvious from
Eq.~\eqref{eq:def-mean-interaction-force} that $f^{\rms int}(x,t)$
vanishes for all $x$, and Eqs.~(\ref{eq:one-particle-equation}) and
\eqref{eq:particle-current-definition} reduce to the equations for
Brownian motion of noninteracting particles subject to the external
force $f^{\rms ext}$.  Accordingly, $\jst(\rho, m\lambda)$,
$m=0,1,\ldots$, is equal to the current $j_0(\rho)$ for
noninteracting particles. This current $j_0(\rho)$ is linearly
dependent on the particle density, $j_0=v_0\rho$, where
\cite{Ambegoakar/Halperin:1969}
\begin{equation}
v_0 = \frac{D\lambda(1-e^{-\beta f \lambda})}
{\int\limits_0^\lambda \dd x \int\limits_x^{x+\lambda} \dd y \exp[\beta(U(y)-fy - U(x) + fx)]}
\label{eq:single-particle-velocity}
\end{equation}
is the mean velocity of a single particle in the steady state.

An alternative way to reason $\jst(\rho, 0)=j_0(\rho)$ is to resort to
the invariance of collective properties under particle exchange for
hardcore interacting point particles \cite{Ryabov/Chvosta:2011}. We
thus refer to this behavior as exchange symmetry effect. It can be
explained by the consideration of path probabilities: To a path
$\mathcal{P}$ in a system $\mathcal{S}$ with hardcore constrains one
can assign the set $\{\mathcal{P}'\}$ of all paths $\mathcal{P}'$ in a
system $\mathcal{S}'$ of independent particles that result from
particle exchanges at all contact points of individual particle
trajectories in $\mathcal{P}$. Because the probability for the set
$\{\mathcal{P}'\}$ in $\mathcal{S}'$ is equal to the probability of
the path $\mathcal{P}$ in $\mathcal{S}$, it follows that averages of
collective quantities, like the current, are equal in $\mathcal{S}'$
and $\mathcal{S}$.

\begin{figure*}[t!]
\includegraphics[width=\textwidth]{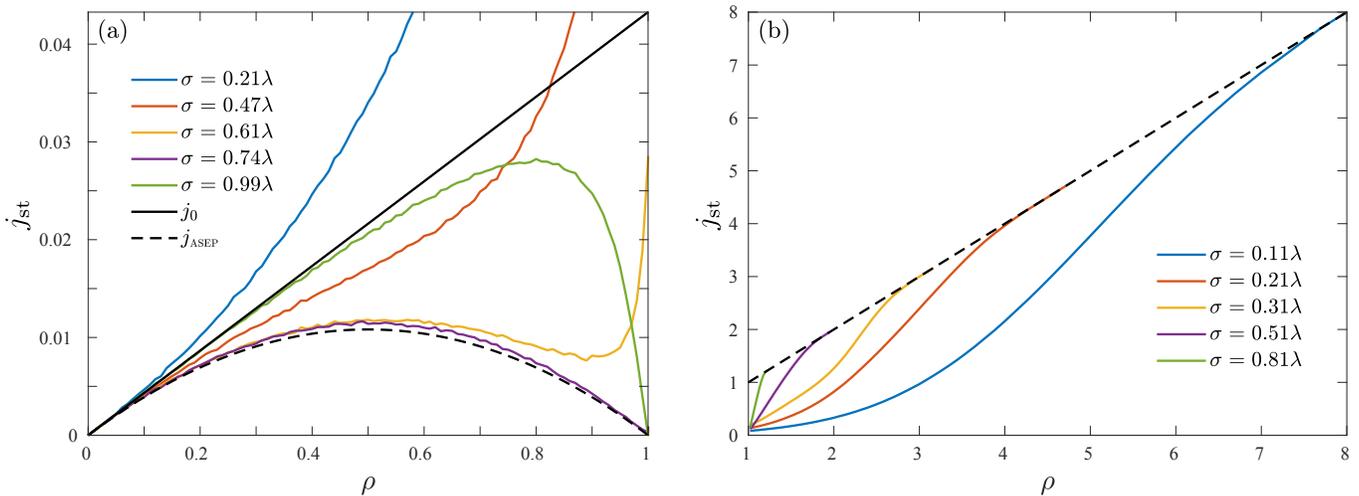}
\caption{Simulated steady-state current in the BASEP with cosine
  potential [Eq.~\eqref{eq:cosine-potential}] as function of the
  density $\rho$ for different particle sizes $\sigma$, (a) for $0
  \leq \rho \leq 1$ and (b) for $1 \leq \rho \leq 8$, [$U_0/(\kB T) =
    6$, $f\lambda/(\kB T)=1$]. In (a) the solid black line marks the
  current of noninteracting particles $j_0(\rho)=v_0\rho$, and the
  dashed line the current-density relation $j_{\rms
    ASEP}(\rho)=v_0\rho(1-\rho)$ of a corresponding ASEP. In (b) the
  dashed black line indicates the upper bound $\mu\rho|f|/\lambda$ of
  the current.}
\label{fig:current-closed}
\end{figure*}

\section{Current in closed system: Specific example}
\label{sec:current_specific_example}
We now turn to a specific implementation and demonstrate the transport
properties and phase transitions in a BASEP with the periodic
potential
\begin{equation}
U(x) = \frac{U_0}{2} \cos \bigg(\frac{2\pi x}{\lambda}\bigg)\,.
\label{eq:cosine-potential}
\end{equation}
As units we choose $\lambda$ for length, $\lambda^2/D$ for time, and
$\kB T$ for energy (and accordingly $\kB T/\lambda$ for
forces). Unless specified otherwise, we set the barrier height
$U_0=6\kB T$ and the drag force $f=1\,\kB T/\lambda$. The barrier
height $U_0 \gg \kB T$ leads to an effective hopping motion between
the potential wells, which resembles the discrete hopping motion in
the lattice ASEP, see the trajectories in Fig.~\ref{fig:model}.

In the simulations of the closed BASEP the system length is set to
$L=100\lambda$ and the time step of the simulation scheme to $\Delta
t=10^{-4} \lambda^2/D$. We checked that our results are not affected
by the finite system length and the chosen time step. The current is
determined by counting the number of particles per unit time crossing
a point of the system in the steady state (and by averaging over many
different points to enhance the accuracy). We performed extensive
Brownian dynamics simulations for a fine grid of $(\sigma,
\rho)$-values covering the area $\{(\sigma, \rho)|\,
0\leq\sigma\leq\lambda, 0 < \rho<\lambda/\sigma\}$. Simulated data
presented in the following figures are always shown as
quasicontinuous lines. When referring to analytical results, this is
explicitly stated.

\subsection{Dependence on particle size and density}
\label{subsec:current-example}
As mentioned in the Introduction, we here give a description of the
current, which includes all particle sizes and also densities $\rho$
in the range $\mathrm{int}(\sigma + 1) < \rho < \lambda/\sigma$,
i.e., regimes not considered in our previous work
\cite{Lips/etal:2018}. Representative examples of current-density
relations for several $\sigma$ are shown in
Fig.~\ref{fig:current-closed} for the density ranges (a) $0 \leq \rho
\leq 1$ and (b) $1 \leq \rho \leq 8$.

In the low density limit $\rho \to 0$, particle interactions become
negligible and the current-density relations approach the linear
relation $j_0(\rho)=v_0\rho$ of noninteracting particles [solid black
  line in Fig.~\ref{fig:current-closed}(a)], where
Eq.~\eqref{eq:single-particle-velocity} yields $v_0=0.043 D/\lambda$
for our parameters. Beyond this common asymptotic limit for all
$\sigma$, the form of the current-density relation varies strongly
with the particle size. For comparison with the lattice model, the
parabolic current-density relations of a corresponding ASEP $j_{\rms
  ASEP}(\rho)=v_0\rho (1-\rho)$ is shown in
Fig.~\ref{fig:current-closed}(a) also (dashed black line).

The change of the current-density relation with the particle size is
due to the interplay of three physical effects \cite{Lips/etal:2018}:
a barrier reduction, a blocking and the exchange-symmetry effect. The
barrier reduction effect occurs if a potential well is occupied by
more than one particle. It leads to a current enhancement compared to
$j_0$, because the particles in the well are pushing each other to
regions of higher potential energy. Thereby the effective barrier for
a transition to the neighboring wells is reduced. In contrast, the
blocking effect lowers the current: if a particle attempts a
transition to the next potential well, its motion can be blocked by a
particle already occupying the neighboring well. The exchange-symmetry
effect emerges as a result of the exact invariance of the current
under particle exchange for commensurable diameters $\sigma=m\lambda$,
as explained Sec.~\ref{subsec:recurrent-dynamics}. It leads to a
continuous deformation of the curved current-density relation for
$\sigma \lesssim m\lambda$ toward the linear dependence for $\sigma =
m\lambda$ [see the curve for $\sigma=0.99\lambda$ in
  Fig.~\ref{fig:current-closed}(a)]. The exchange symmetry effect thus
becomes already notable for $\sigma \lesssim \lambda$.

To understand how these effects influence $\jst(\rho, \sigma)$ in
Fig.~\ref{fig:current-closed}(a), we discuss the curves with
increasing particle size.  For small $\sigma$, $\jst(\rho, \sigma)$ is
monotonically increasing with $\rho$ and always larger than
$j_0(\rho)$, see the curve for $\sigma=0.21\lambda$ in
Fig.~\ref{fig:current-closed}(a). The enhancement compared to
$j_0(\rho)$ is due to the barrier reduction effect, which prevails for
small $\sigma$ because of a high multi-occupation probability of
potential wells.  With increasing $\sigma$, the influence of the
blocking effect becomes stronger, which leads to currents smaller than
$j_0(\rho)$ at intermediate $\sigma$ and not too high $\rho$. In this
regime, $\jst(\rho, \sigma)$ is still monotonically increasing with
$\rho$, see the curve for $\sigma=0.47\lambda$ in
Fig.~\ref{fig:current-closed} (a). The strong rise of $\jst(\rho,
\sigma)$ at larger $\rho$ values is caused by double occupancies that
are propagating through clusters of single occupied wells in a
cascade like manner \cite{Lips/etal:2018, Ryabov/etal:2019}.

Beyond a certain particle size $\sigma_{\rms c}$, a local maximum and
local minimum appears in $\jst(\rho, \sigma)$ at densities $\rho_{\rms
  max} = \rho_{\rms max} (\sigma)$ and $\rho_{\rms min} = \rho_{\rms
  min}(\sigma)$, see the curves for $\sigma=0.61\lambda$ in
Fig.~\ref{fig:current-closed}(a). When further enlarging $\sigma$, the
blocking effect dominates the behavior for all $0 \leq \rho \leq 1$
and the current-density relations approach $j_{\rms ASEP} (\rho)$ as a
limiting curve with a maximum at $\rho \simeq 0.5$. This occurs in the
range $0.74\lambda \lesssim \sigma \lesssim 0.82\lambda$ for our
setup. Close to the commensurate diameter $\sigma = \lambda$, the
exchange symmetry effect becomes relevant. As a consequence, the
position of the maximum in $\jst(\rho, \sigma)$ moves to higher
densities and the current approaches $j_0(\rho)$ from below.

If the number of particles exceeds the number of potential wells,
i.e., when $\rho > 1$, then the particles cannot all be localized close to
a potential minimum and double or multi-occupied wells are permanently
present. This leads to a strong increase of the particle current with
$\rho$ toward values much larger than those seen in
Fig.~\ref{fig:current-closed}(a) for $\rho < 1$ [note the different
  scales of the axes for the current in
  Figs.~\ref{fig:current-closed}(a) and \ref{fig:current-closed}(b)]. The upper bound $\mu
\rho f/\lambda$ (see Sec.~\ref{subsec:current-bounds}) is shown as the
dashed black line in Fig.~\ref{fig:current-closed} (b). In the limit
$\rho \rightarrow \lambda/\sigma$ of complete filling, the curves
approach the upper bound from below.

\begin{figure}[t!]
\includegraphics[width=\columnwidth]{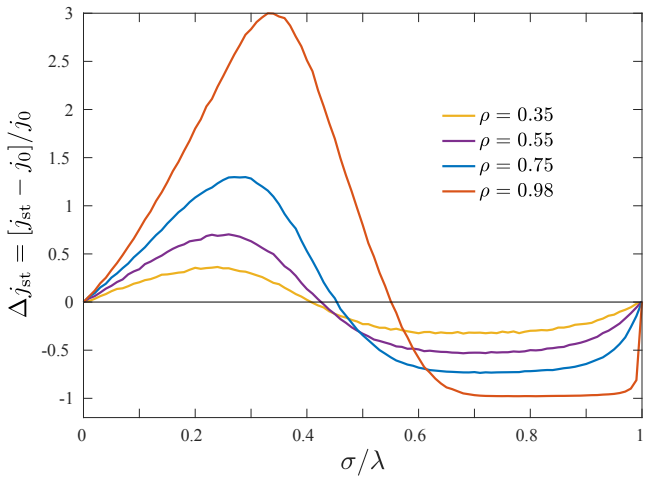}
\caption{Change of the steady-state current in the
    BASEP with cosine potential relative to that of independent
  particles as a function of the particles diameter for several fixed
  densities [$U_0/(\kB T) = 6$, $f\lambda/(\kB T)=1$].}
\label{fig:sigma_dependence}
\end{figure}

How the influence of the barrier reduction, blocking and exchange
symmetry effects changes with the particle size becomes particularly
transparent when plotting the relative change $\Delta \jst(\rho,
\sigma) = [\jst(\rho,\sigma) - j_0(\rho)]/j_0(\rho)$ of the current
with respect to $j_0(\rho)$ as a function of $\sigma$ for different
fixed $\rho$. Corresponding curves shown in
Fig.~\ref{fig:sigma_dependence} all display a local maximum at a value
$\sigma_{\rms max}(\rho)$ and show a plateau like behavior in an
intermediate $\sigma$ range. For $\sigma$ below the onset of the
plateau like regime, the barrier reduction and the blocking effect
compete with each other, where the barrier reduction governs the
change of $\Delta\jst$ for $\sigma < \sigma_{\rms max}$ and the
blocking effect for $\sigma > \sigma_{\rms max}$. In the plateau like
regime, the barrier reduction effect becomes almost negligible. For
$\sigma \to \lambda$ the exchange symmetry effect causes the curves to
increase back to $j_0(\rho)$, which is reached at $\sigma=\lambda$.

\begin{figure}[t!]
\includegraphics[width=\columnwidth]{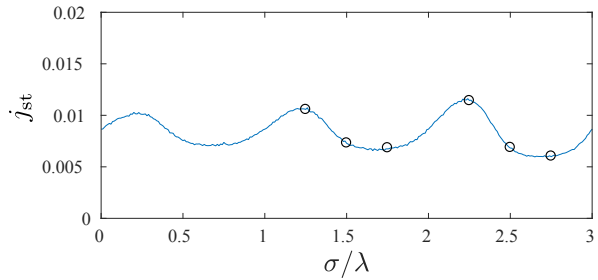}
\caption{Steady-state current as a function of
  $\sigma/\lambda$ at a fixed density $\rho=0.2$ [$U_0/(\kB T) = 6$,
    $f\lambda/(\kB T)=1$]. The blue line is obtained from the
  simulated data for $0 \leq \sigma \leq \lambda$ and $0 \leq \rho
  \leq \lambda/\sigma$ and by applying
  Eq.~\eqref{eq:particle-diameter-mapping-b} for $\sigma>\lambda$.
  The circles indicate results from simulations of systems with
  diameters $\sigma/\lambda=\{1.25,1.50,1.75,2.25,2.50, 2.75\}$.
\label{fig:verification-scaling-relation}}
\end{figure}

Having described the current for all densities $0 \leq \rho \leq
\lambda/\sigma$ in the range of particle sizes $0 \leq \sigma \leq
\lambda$, we are able to use the mapping in
Eq.~\eqref{eq:particle-diameter-mapping-b} to derive
$\jst(\rho,\sigma)$ for diameters $\sigma>\lambda$ at arbitrary
densities. This is exemplified in
Fig.~\ref{fig:verification-scaling-relation}, where we show
$\jst(\rho,\sigma)$ as function of $\sigma$ for a fixed density
$\rho=0.2$. The curve (blue line) for $\sigma>\lambda$ was calculated
from the simulated data in the range $0 \leq \sigma \leq \lambda$ and
$0 \leq \rho \leq \lambda/\sigma$ by applying
Eq.~\eqref{eq:particle-diameter-mapping-b}. To demonstrate the
validity of this equation, we performed additional simulations for
certain $\sigma > \lambda$ for which the current (black circles) is
shown in Fig.~\ref{fig:verification-scaling-relation}.

\begin{figure}[t!]
\includegraphics[width=\columnwidth]{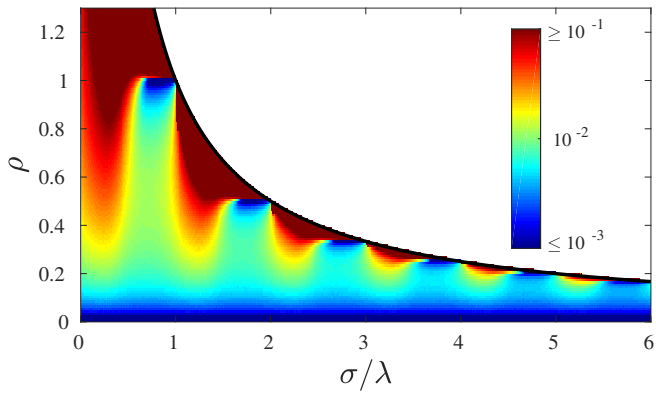}
\caption{Color-coded representation of $\jst(\rho,\sigma)$ in the
  $\sigma$-$\rho$--plane [$U_0/(\kB T) = 6$, $f\lambda/(\kB
    T)=1$]. Values below $10^{-3}D/\lambda^2$ and above
  $10^{-1}D/\lambda^2$ are indicated in dark blue and red,
  respectively. The scale bar specifies the color coding for the other
  values. The displayed current for diameters $\sigma
    >\lambda$ was obtained from the simulated data for $0 \leq \sigma
    \leq \lambda$ using the
    mapping~\eqref{eq:particle-diameter-mapping-b}. The
    mapping~\eqref{eq:particle-diameter-mapping-b} was independently
    verified for several values of $\sigma$, $\sigma >\lambda$ by
    Monte Carlo simulations; a particular example is shown in
    Fig.~\ref{fig:verification-scaling-relation}.}
\label{fig:recurrent_dynamics}
\end{figure}

The emerging recurrence pattern in the $\sigma$-$\rho$--plane is shown
in Fig.~\ref{fig:recurrent_dynamics}, where values of the current are
represented by a color coding. Note that this pattern is not periodic
in $\sigma$ (see also Fig.~\ref{fig:verification-scaling-relation}),
but has a more complicated structure because of the necessary
rescaling of the density.

When comparing the behavior of the steady-state current in the BASEP
and ASEP, the blocking effect is present in both models. In contrast,
the barrier reduction effect has no analog in the lattice model,
because multi-occupation of a site is forbidden in the standard
ASEP. The exchange-symmetry effect is also absent in lattice models,
even if one introduces a generalized $l$-ASEP, where the particles
occupy $l$ lattice sites. This is because the particle size is an
integer multiple of the lattice constant and accordingly there is no
continuous transition toward a commensurate diameter.

Nevertheless, the reasoning in Sec.~\ref{subsec:recurrent-dynamics}
can be taken over to the $l$-ASEP with $l>1$, $N$ particles, $M$
lattice sites with periodic boundary conditions, and hopping rates
$\Gamma_+$ and $\Gamma_-$ in and against bias direction. A
transformation $l \to 1$ and $M \to M - N(l-1)$ corresponds to the
transformation considered in Sec.~\ref{subsec:recurrent-dynamics} for
the BASEP with $m=(l-1)$. Hence from
Eq.~\eqref{eq:particle-diameter-mapping-b} and with $j_{\rms
  ASEP}(\rho) = (\Gamma_+ - \Gamma_-) \rho (1-\rho)$ we obtain
\begin{align}
j_{l{\rms -ASEP}}(\rho) &= [1-(l-1)\rho]\,
j_{\rms ASEP}\left(\frac{\rho}{1-(l-1)\rho}\right)\nonumber\\
&= (\Gamma_+ - \Gamma_-) \frac{\rho (1 - l\rho)}{1- (l-1)\rho} \, .
\end{align}
The reduction from an $l$-ASEP to the standard ($l=1$) ASEP has been
used in the literature before to obtain the current-density relation
\cite{Schoenherr/Schuetz:2004}.  In contrast to the equality of the
current in the BASEP for commensurate $\sigma=m\lambda$,
$m=0,1,2,\ldots$, the currents in the $l$-ASEP change with $l$. This
is because $j_{\rms ASEP}(\rho)$ for the smallest length $l=1$ in the
$l$-ASEP is a nonlinear function of $\rho$, while $j_0(\rho)$ for the
smallest length $\sigma=0$ (point particles) in the BASEP varies
linearly with $\rho$.

\subsection{Temperature dependence}
\label{subsec:temperature-dependence} 
With decreasing temperature, the particles become stronger localized
at the minima of the potential wells. In the $\sigma$ regime of
dominant blocking effect, the current-density relation therefore
follows more closely $j_{\rms ASEP}$. This causes the plateau like
regimes discussed in connection with Fig.~\ref{fig:sigma_dependence}
to extend and to become extremely flat in the zero temperature
(low-noise) limit.

\begin{figure}
\includegraphics[width=\columnwidth]{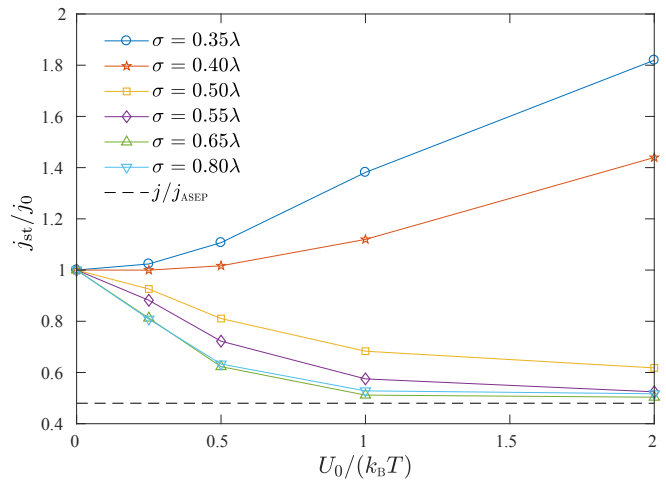}
\caption{Ratio $\jst/j_0$ as a function of $U_0/(\kB T)$ at fixed
  $\rho=0.52$ and $f\lambda/U_0 = 1/6$ for different particle
  diameters. The dashed line marks the the low-temperature limit
  $j_{\rms ASEP}/j_0$ in the regime of dominating blocking effect.}
\label{fig:temperature-dependence}
\end{figure}

However, the stronger localization at the minima of the potential does
not mean that the barrier reduction effect disappears for small
$\sigma$. In relation to the current $j_0$ of a single particle,
double occupancies of wells lead to an enhancement at arbitrary low
temperatures. As the barrier reduction should become almost
independent of temperature at low $T$, $\jst/j_0$ will even be larger
for lower temperatures. Hence, when considering the dependence of
$\jst/j_0$ on temperature for decreasing $T$, we expect a decrease for
large $\sigma$ where the blocking effect prevails, and an increase for
small $\sigma$ where the barrier reduction dominates. This is indeed
the case and demonstrated in Fig.~\ref{fig:temperature-dependence},
which shows $\jst/j_0$ as a function of $U_0/(\kB T)$ for different
$\sigma$ at a fixed density $\rho=0.52$ (and the same ratio
$f\lambda/U_0 = 1/6$ as considered before).

\subsection{Analytical approaches in the limit of small driving}
\label{subsec:small-driving}
Using the exact expression for the steady-state current in
Eq.~\eqref{eq:ness-current}, we can calculate an approximation for
small drag forces. In an equilibrium system ($f=0$), $\jst=0$ and
$\overline{f_{\rm st}^{\rms int}}=0$. In the linear-response regime,
we thus obtain from Eq.~\eqref{eq:ness-current}
\begin{equation}
\jst(\rho, \sigma) \sim 
\frac{\mu(1+\alpha)}{\displaystyle\frac{1}{\lambda}
\int_0^\lambda \dd x \frac{1}{\varrho_{\rms eq}(x)}}f\,, 
\quad \alpha = \frac{\overline{f^{\rms int}_{\rm st}}}{\partial f}\bigg|_{f=0}\,,
\label{eq:linear-response}
\end{equation}
where $\varrho_{\rms eq}(x)$ is the local density profile in
equilibrium. The equilibrium density profile (in the grand-canonical
ensemble) can be obtained by minimizing the exact density functional
\begin{align}
\Omega[\varrho(x)] = &\int\limits_0^\lambda \dd x \varrho(x) \biggl\{
U(x) - \mu_{\rms ch} \nonumber \\
&{}- \kB T \left[ 1 - \ln\left(
\frac{\varrho(x)}{1 - \eta(x)} \right) \right]\biggr\}
\label{eq:percus}
\end{align}
for hard rods in one dimension \cite{Percus:1976}. Here, 
$\mu_{\rms ch}$ is the chemical potential and
\begin{equation}
\eta(x) = \int\limits_{x-\sigma}^x \dd y\, \varrho(y) \, .
\label{eq:eta}
\end{equation}
Minimizing $\Omega[\varrho(x)]$ in Eq.\eqref{eq:percus} yields
\begin{align}
0=\frac{\delta \Omega[\varrho]}{\delta \varrho}
\bigg|_{\varrho=\varrho_{\rms eq}} &= \ln\left[
\frac{\varrho_{\rms eq}(x)}{1-\eta_{\rms
    eq}(x)}\right] \label{eq:structure-equation}\\ 
&+\int\limits_{x}^{x+\sigma} \dd y \frac{\varrho_{\rms
    eq}(y)}{1-\eta_{\rms eq}(y)}
+ \beta [U(x) - \mu_{\rms ch}] \, .\nonumber
\end{align}
We discretized this equation and solved it numerically under periodic
boundary conditions [$\varrho_{\rms eq}(x)=\varrho_{\rms
    eq}(x+\lambda)$] to obtain the density profile for a given
chemical potential $\mu_{\rms ch}$ (or density $\rho$). Inserting the
solution for $\varrho_{\rms eq}(x)$ in Eq.~\eqref{eq:linear-response}
and setting $\alpha=0$ in Eq.~\eqref{eq:linear-response}, we obtain a
small-driving approximation (SDA) for $\jst(\rho,\sigma)$.

\begin{figure}
\includegraphics[width=\columnwidth]{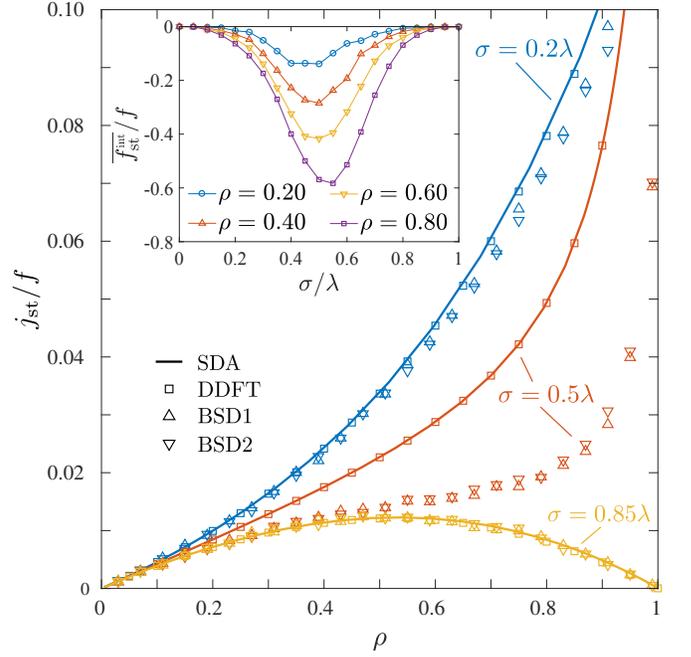}
\caption{Current-density relations for different particle diameters
  and $U_0/(\kB T) = 6$, $f\lambda/(\kB T)=0.2$ obtained from Brownian
  dynamics simulations with algorithms from
  Ref.~\cite{Behringer/Eichhorn:2012} (BDS1) and
  Ref.~\cite{Scala:2012} (BDS2), the SDA
  [Eq.~\eqref{eq:linear-response} with $\alpha=0$], and the DDFT. The inset shows the
  period-averaged mean interaction force $\overline{f^{\rms int}_{\rm
      st}}$ in dependence of the particle diameter for different
  $\rho$. It has the largest magnitude close to $\sigma=0.5\lambda$,
  where SDA and DDFT results quantitatively differ from the ones from
  the Brownian dynamics simulations because these theories
  underestimate the magnitude of~$\overline{f^{\rms int}_{\rms st}}$.}
\label{fig:linear-response}
\end{figure}

Keeping nonzero $\alpha$ requires more advanced approaches. A widely
used method is the dynamical density functional theory (DDFT)
\cite{Marconi/Tarazona:1999, Marconi/Tarazona:2000}, where the
two-particle density at contact in
Eq.~\eqref{eq:def-mean-interaction-force} is related to the
single-particle density as in an equilibrium system:
\begin{subequations}
\begin{align}
\varrho^{(2)}(x,x+\sigma,t)&=\frac{\varrho(x,t)\varrho(x+\sigma,t)}{1-\eta(x+\sigma,t)}\,,\\
\varrho^{(2)}(x,x-\sigma,t) &=\frac{\varrho(x,t) \varrho(x-\sigma,t)}{1-\eta(x, t)}\,.
\end{align}
\label{eq:ddft-two-particle-density}
\end{subequations}
Here, $\eta(x,t)$ is given by Eq.~\eqref{eq:eta} with $\varrho(y)$
replaced by $\varrho(y,t)$. Combining
Eqs.~\eqref{eq:one-particle-equation},
\eqref{eq:particle-current-definition},
\eqref{eq:def-mean-interaction-force} and
\eqref{eq:ddft-two-particle-density} results in a nonlinear and
nonlocal evolution equation for the density:
\begin{align}
\frac{\partial \varrho(x,t)}{\partial t} &= \frac{\partial}{\partial
  x}
\bigg\{D\frac{\partial\varrho(x,t)}{\partial x}
-\mu f^{\rms ext}(x) \varrho(x, t)  \label{eq:ddft-equation}\\ 
&\hspace{1em}{}- D \varrho(x,t ) \bigg[  
\frac{\varrho(x-\sigma, t)}{1-\eta(x, t)} - \frac{\varrho(x+\sigma, t)}{1-\eta(x+\sigma,t)} 
\bigg] 
\bigg\} \nonumber.
\end{align}
To find its stationary solution we used two methods. First, we
propagated an initial density profile into the stationary regime with
a forward-time central-space scheme. Secondly, we solved the
corresponding stationary equation ($\partial \varrho/\partial t = 0$)
with an iterative scheme. Both methods lead to equal results.

Current-density relations obtained with the SDA, DDFT and Brownian
dynamics simulations with the algorithms given in
Ref.~\cite{Behringer/Eichhorn:2012} (BDS1) and Ref.~\cite{Scala:2012}
(BDS2) are shown in Fig.~\ref{fig:linear-response} for three particle
diameters $\sigma=0.2\lambda$ (blue), $\sigma=0.5\lambda$ (orange),
and $\sigma=0.85 \lambda$ (yellow) in the small bias regime at $f=0.2
\kB T/ \lambda$.  We see that the DDFT results are almost
indistinguishable from the SDA results for all shown diameters, and
the results from BDS1 and BDS2 are in very good agreement. Overall the
SDA and DDFT capture the qualitative features of the current-density
relations.

\begin{figure}[t!]
\includegraphics[width=\columnwidth]{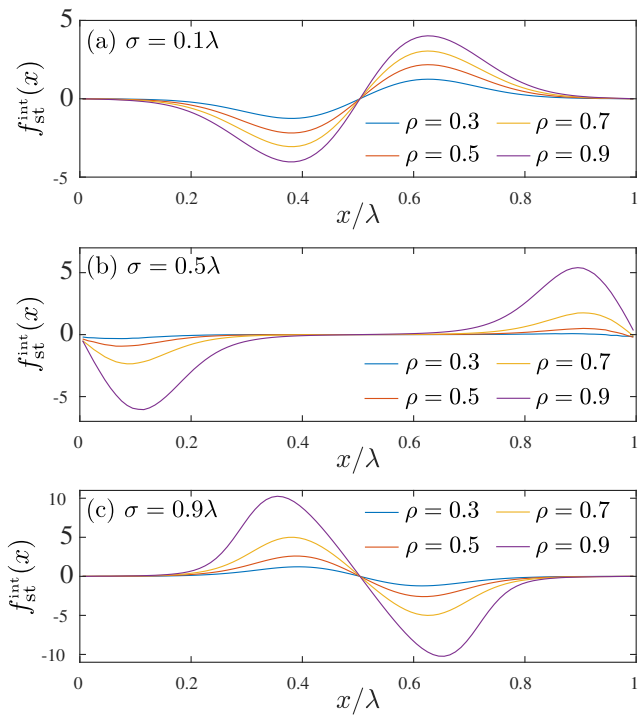}
\caption{Profiles of the local mean interaction force in the steady
  state for different densities and particle diameters (a)
  $\sigma=0.1\lambda$, (b) $\sigma=0.5\lambda$, and (c)
  $\sigma=0.0\lambda$ [$U_0/(\kB T) = 6$, $f\lambda/(\kB T)=0.2$].}
\label{fig:mean-interaction-force}
\end{figure}

However, comparing the results of the SDA and DDFT with that of the
simulations quantitatively, we observe deviations that become largest
for intermediate particle sizes close to $\sigma=0.5\lambda$. The
reason is that both the SDA and DDFT underestimate the magnitude of
the period-averaged mean interaction force $\overline{f_{\rm st}^{\rms
    int}}$, which we determined from BDS1. To get insight into its
behavior, we show in the inset of Fig.~\ref{fig:linear-response}
$\overline{f_{\rm st}^{\rms int}}$ as a function of $\sigma/\lambda$
for different densities. As can be seen from the data,
$\overline{f_{\rm st}^{\rms int}}$ is always opposite to the bias $f$,
in agreement with our discussion in
Sec.~\ref{subsec:current-bounds}. Its magnitude is largest close to
$\sigma = 0.5 \lambda$.  The corresponding minimum shifts slightly to
higher $\sigma$ and becomes more pronounced with increasing $\rho$.

Closer inspection shows that the local mean interaction force $f_{\rm
  st}^{\rms int}(x)$ in the steady state can be both parallel and
antiparallel to the drag force $f$ and that it is always small close
to the local extrema of the external potential, where $f^{\rms
  ext}(x)\simeq0$. This is demonstrated in
Fig.~\ref{fig:mean-interaction-force}, where we show representative
profiles $f_{\rm st}^{\rms int}(x)$ for different $\rho$ and $\sigma$.
The shape of these profiles changes significantly with the particle
diameter.  For small $\sigma=0.1\lambda$
[Fig.~\ref{fig:mean-interaction-force}(a)], a region of negative and
positive $f_{\rm st}^{\rms int}(x)$ occurs next to the minimum of the
external potential in ($x\lesssim 0.5\lambda$) and against bias
($x\gtrsim0.5\lambda$) direction, respectively. At an intermediate
$\sigma=0.5\lambda$ [Fig.~\ref{fig:mean-interaction-force}(b)], these
regions of negative and positive $f_{\rm st}^{\rms int}(x)$ have
shifted to locations close to the potential barriers, and an extended
regime of negligible $f_{\rm st}^{\rms int}(x)\simeq0$ appears around
$x=0.5\lambda$. At large $\sigma=0.9\lambda$
[Fig.~\ref{fig:mean-interaction-force}(c)], the profile from
Fig.~\ref{fig:mean-interaction-force}(a) appears to be kind of
inverted, with now a region of positive and negative $f_{\rm st}^{\rms
  int}(x)$ occurring for $x\lesssim 0.5\lambda$ and $x\gtrsim 0.5\lambda$,
respectively.  As for the density dependence, it changes the magnitude
of $f_{\rm st}^{\rms int}(x)$ along with a shift of the positions of
its local minima and maxima.

These changes of the profiles $f_{\rm st}^{\rms int}(x)$ can be
understood by noting that for small $\sigma$, a particle located at a
position left (right) of the potential minimum collides with other
particles in multiple-occupied wells, which more frequently are coming
from the right (left). This leads to a mean repulsive force that tends
to push the particle further away from the potential minimum, i.e., we
obtain $f_{\rm st}^{\rms int}(x)<0$ for $x\lesssim0.5\lambda$ and $f_{\rm
  st}^{\rms int}(x)>0$ for $x\gtrsim0.5\lambda$.  At intermediate $\sigma$,
multiple occupancies become unlikely and the particles preferentially
occupy positions close to the potential minima.  A particle positioned
near a potential minimum rarely collides with other particles so that
$f_{\rm st}^{\rms int}(x)\simeq0$ in a region around the minimum.
Particles located close to the potential barriers now interact most
strongly with other particles. For large $\sigma$, the blocking effect
pushes the particles toward their minima, i.e., we find $f_{\rm
  st}^{\rms int}(x)>0$ for $x\lesssim0.5\lambda$ and $f_{\rm st}^{\rms
  int}(x)<0$ for $x\gtrsim0.5\lambda$.

\section{Phase transitions in open system}
\label{sec:open-basep}

\subsection{Extremal current principles}
\label{subsec:phasetrans}
A striking feature of the ASEP is the occurrence of nonequilibrium
phase transitions in open systems coupled to particle reservoirs
\cite{Krug:1991, Schuetz/Domany:1993}. These phase transitions
manifest themselves as discontinuous changes of the bulk density in
dependence of control parameters specifying the coupling to the
reservoirs, or by jumps in the derivatives of the bulk densities with
respect to these control parameters. How theses phases change with
experimentally tunable control parameters depends on details of the
system reservoir couplings \cite{Dierl/etal:2012,
  Dierl/etal:2013}. However, all possible phases can be derived from
extremal current principles \cite{Krug:1991, Antal/Schuetz:2000,
  Dierl/etal:2013}. The arguments leading to these principles are
quite general for driven diffusive systems and are valid also for
driven Brownian motion \cite{Maass/etal:2018}.

We focus here on an open system with $M$ potential wells $i=1,\ldots,
M$ coupled to reservoirs L and R at its left ($i=1$) and right ($i=M$)
end. The ``filling factor'' of the $i$th well in the steady state is
given by (note that $\varrhost(x)$ is not a periodic function in the
open system)
\begin{equation}
\rho_i=\int_{(i-1)\lambda}^{i\lambda}\dd x
\,\varrhost(x), \quad i=1,\ldots,M\,,
\label{eq:period-average-profile}
\end{equation}
and we refer to this as period-averaged density profile.

In the thermodynamic limit ($M\to \infty$), this profile $\rho_i$
exhibits a constant bulk value $\rhoB$ in the interior of the system,
which defines the order parameter of the phase transitions. For finite
large $M$, $\rhoB$ refers to the density of an extended plateau like
region in the interior of the system. Using the extremal current
principles, $\rhoB$ is given by
\begin{equation}
\rhoB=\left\{\begin{array}{l@{\hspace{1em}}l}
\displaystyle\argmin_{\rhoL\le\rho\le\rhoR}\{\jst(\rho,\sigma)\}\,, & \rhoL\le\rhoR\,,\\[3ex]
\displaystyle\argmax_{\rhoR\le\rho\le\rhoL}\{\jst(\rho,\sigma)\}\,, & \rhoR\le\rhoL\,.
\end{array}\right.
\label{eq:min-max-cur}
\end{equation}
Here $\rhoL$ and $\rhoR$ can be any densities bounding a monotonically
varying region encompassing the plateau part from the left and right
side. For globally monotonic profiles in particular, it is possible to
interpret $\rhoL$ and $\rhoR$ as reservoir densities.

In general, however, interaction effects lead to density oscillations
close to the boundaries, and only a specific coupling called
``bulk-adapted" ensures a global monotonic behavior and a controlled
generation of all possible phases \cite{Dierl/etal:2012,
  Dierl/etal:2013}. While such bulk-adapted coupling can be realized
in a systematic manner in lattice models, its implementation in
Brownian dynamics of interacting particles remains challenging. We
will take a pragmatic approach and present in
Sec.~\ref{subsec:simulated_phases} a simple coupling scheme of
particle injection and ejection for the simulation of the phases.

\begin{figure}[t!]
\includegraphics[width=\columnwidth]{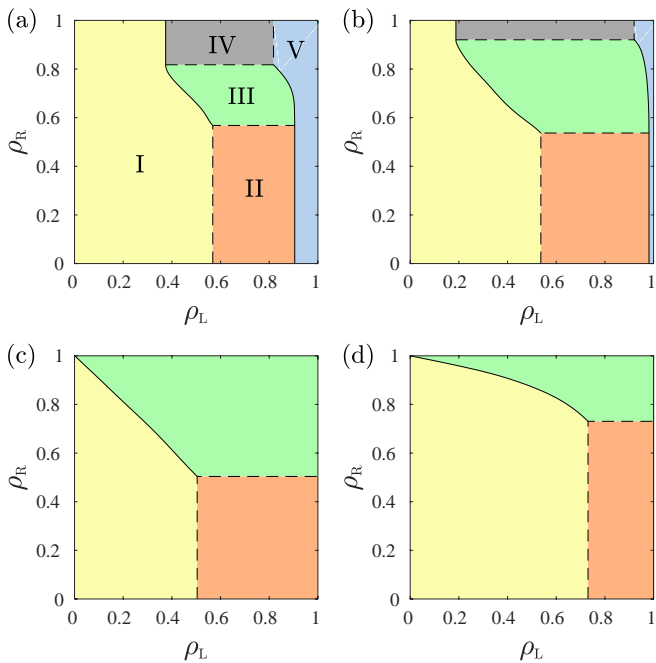}
\caption{Phase diagrams of the open BASEP predicted by applying the
  extremal current principles to the bulk current-density relation for
  $U_0/(\kB T) = 6$ and $f\lambda/(\kB T)=1$. In (a) the particle size
  is $\sigma=0.58\lambda$, and in (b) $\sigma=0.62\lambda$, (c)
  $\sigma=0.75\lambda$, and (d) $\sigma=0.98\lambda$. The phases I and
  V are left-boundary induced phases, phase II is a maximal current
  phase, phases III is a right-boundary induced phase, and phase IV is
  a minimal current phase. The phases are equally colored in all
  panels. Solid lines mark first order phase transitions and dashed
  lines second order phase transitions.}
\label{fig:phase-diagrams}
\end{figure}

\subsection{Phases derived from extremal current principles}
\label{subsec:phases_from_principles}
Application of the extremal current principles \eqref{eq:min-max-cur}
to the simulated current-density relations of the closed system yields
the phase diagrams for the open system. The change of the shape
$\jst(\rho,\sigma)$ with the particle diameter leads to different
types of diagrams. Representative examples are shown in
Figs.~\ref{fig:phase-diagrams}(a)-\ref{fig:phase-diagrams}(d). For $\sigma<\sigma_{\rm c}$, no
phase transitions occur, because the current as a function of density
exhibits no local extrema.  For $\sigma \gtrsim \sigma_{\rm c}$, the
maximum number of five phases appears, which we labeled I-V in
Fig.~\ref{fig:phase-diagrams}(a). These phases are colored equally in
all other panels. Phases I, III and V are boundary-matching with
$\rhoB=\rhoL$ in phases I and V, and $\rhoB=\rhoR$ in phase III.  Phases
II and IV are maximal and minimal current phases with
$\rhoB=\rho_{\rms max}$ and $\rhoB=\rho_{\rms min}$, respectively
($\rho_{\rms max}$ and $\rho_{\rms min}$ are the densities at which
the current-density relation has a local maximum and minimum, see
Sec.~\ref{subsec:current-example}).  Solid lines mark first-order
transitions, where $\rhoB$ changes discontinuously when the line is
crossed, and dashed lines mark second-order transition, where $\rhoB$
varies continuously but the gradient of $\rhoB$ with respect to
$\rhoL$ and $\rhoR$ changes discontinuously.

With increasing $\sigma>\sigma_{\rm c}$, the phases IV and V shrink
[Fig.~\ref{fig:phase-diagrams}(b)].  This shrinkage continues up to
the point where the phase diagram is similar to the one for a
corresponding ASEP with three phases
[Fig.~\ref{fig:phase-diagrams}(c)]. Close to the commensurate
diameter, the diagram with the three phases becomes asymmetric due to
the change of $\rho_{\rms max}(\sigma)$
[Fig.~\ref{fig:phase-diagrams}(d)].

\begin{figure}[t!]
\includegraphics[width=\columnwidth]{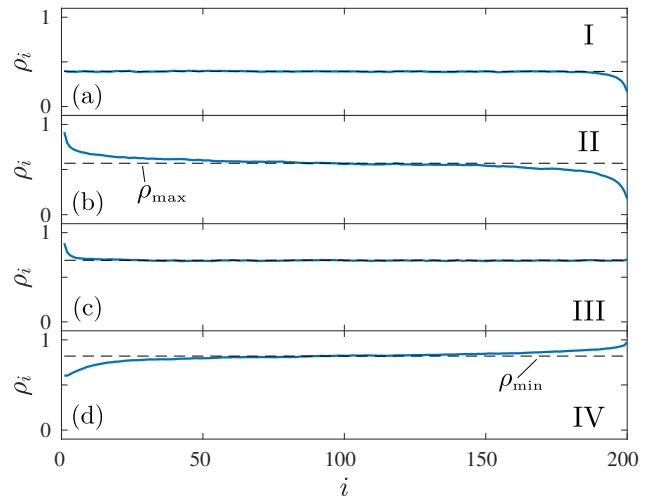}
\caption{Period-averaged density profiles in the different
  nonequilibrium phases of the open system for $\sigma=0.58\lambda$
  [$U_0/(\kB T) = 6$, $f\lambda/(\kB T)=1$]: (a) boundary-matching
  phase I with $\rhoB\cong\rhoL=0.39$, (b) maximal current phase II
  with $\rhoB\cong\rho_{\rms max}=0.57$, (c) boundary-matching phase
  III with $\rhoB\cong\rhoR=0.69$, and (d) minimal current phase IV
  with $\rhoB\cong\rho_{\rms min}=0.82$. Dashed lines in (a)-(d)
  correspond to the bulk value in the center of the system (see
  text).}
 \label{fig:phase-examples}
 \end{figure}

\subsection{Simulated phases}
\label{subsec:simulated_phases}
To verify the phase diagrams derived from the extremal current
principles in the Brownian dynamics simulations, we consider particles
of size $\sigma=0.58\lambda$, where Fig.~\ref{fig:phase-diagrams}(a)
gives the predicted phase diagram.  For the exchange of particles with
the reservoirs, we used the following method: If the leftmost
(rightmost) potential well is empty, then a particle is injected with a
rate $\alpha_{\rms L}$ ($\alpha_{\rms R}$). Injected particles are
placed at the distance $(\lambda-\sigma)$ away from the boundary. This
guarantees that no particle overlaps occur. Ejection of a particle to
a reservoir is implemented by removing it from the system once its
center position crosses the left or right boundary.
 
In a simulation with fixed ($\alpha_{\rms L}$, $\alpha_{\rms R}$) the
system approaches a steady state with density profile $\rho_i$
[Eq.~\eqref{eq:period-average-profile}] after an initial transient
time. Examples of simulated density profiles for the nonequilibrium
phases I-IV are shown in Fig.~\ref{fig:phase-examples}(a)-\ref{fig:phase-examples}(d). The
bulk density is extracted from an average of $\rho_i$ around the
center of the system, $\rhoB = \sum_{i=L/2-d}^{L/2+d} \rho_i /(2d)$
with $d=5$. These bulk values can be compared with the predicted
ones. In all Figs.~\ref{fig:phase-examples}(a)-\ref{fig:phase-examples}(d) we obtain a very
good agreement.

When varying the injection rates $\alpha_{\rms L}$ and $\alpha_{\rms
  R}$, the period-averaged boundary densities in the leftmost and
rightmost potential well change. Hence, each simulation run with given
$(\alpha_{\rms L}, \alpha_{\rms R})$ results in one set of
period-averaged boundary and bulk densities $(\rhoL, \rhoR, \rhoB)$ in
the nonequilibrium steady state
\footnote{In most simulations, the period-averaged density profile
  $\rho_i$ turned out to be monotonically varying with $i$. Only for
  very high $\alpha_{\rms L}$ and $\alpha_{\rms R}$ small dips appear
  close to the system boundaries. Then $\rhoL$ and $\rhoR$ were
  determined from the region of monotonically varying profile.}. By
performing many simulation runs for different $(\alpha_{\rms L},
\alpha_{\rms R})$, the various phases can be identified.

\begin{figure}
\includegraphics[width=\columnwidth]{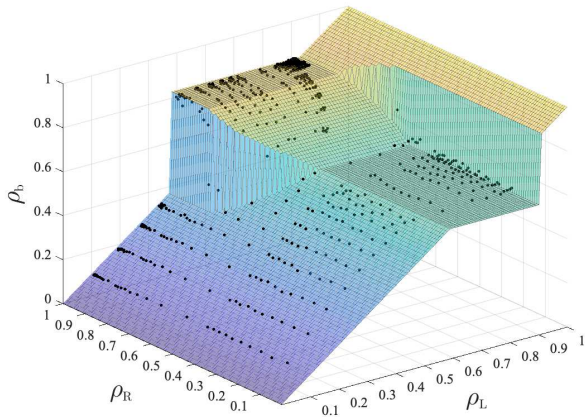}
\caption{Bulk density $\rhoB$ as function of $\rhoL$ and $\rhoR$ for
  particle size $\sigma=0.58\lambda$ [$U_0/(\kB T)=6$, $f\lambda/(\kB
    T)=1$]. The colored surface represents the prediction from the
  extremal current principles in Eq.~\eqref{eq:min-max-cur} and the
  points were extract from simulated density profiles for
  $L=200\lambda$ (as shown for the examples given in
  Fig.~\ref{fig:phase-examples}).}
\label{fig:phase-diagram-simulation}
\end{figure}

Correspondingly simulated data points in a system of length
$L=200\lambda$ are shown in Fig.~\ref{fig:phase-diagram-simulation}
together with the surface $\rhoB=\rhoB(\rhoL,\rhoR)$ obtained from the
extremal current principles~\eqref{eq:min-max-cur}.  As can be seen
from the figure, the simulated values for the bulk density (black
points) lie very closely to the predicted surface
$\rhoB(\rhoL,\rhoR)$. In particular, the discontinuous first-order
transitions from the left-boundary induced phase I to the minimal
current phase IV and to the right-boundary induced phase III are
clearly visible, as well as the continuous transitions between phase I
and the maximal current phase II.  Also the continuous transitions
between phases III and IV, and between phases II and III can be seen
in the simulated data. As for phase V, our simple injection and
ejection method did not generate the very high boundary densities
$\rhoL$ in this phase.
  
\section{Conclusions}
\label{sec:conclusions}
The BASEP represents a wide class of simple models for
Brownian single-file transport in periodic structures, designed to
explore and understand basic physical mechanisms of nonequilibrium driven
motion. We expect that the competition between the barrier reduction,
blocking and exchange symmetry effect plays a 
decisive role in systems with all kinds of
short-ranged interactions. The BASEP can thus serve
as a reference for more complicated nonequilibrium systems, similar as
the hard-sphere fluid became a useful basis in equilibrium liquid
theory. One remaining challenge is to develop better analytical
approaches for the local mean interaction force acting on a
particle. With that at hand, perturbative treatments for interactions
beyond hardcore exclusion could be developed.

We have focused in this work on a sinusoidal form of the external
potential. In the approximate analytical treatment, arbitrary forms of
the periodic potential $U(x)$ can be used by inserting it in the
density functional in Eq.~\eqref{eq:percus}. Based on our findings for
the sinusoidal potential, we conjecture that this method will capture
the transport behavior on a qualitative level. A better quantitative
agreement requires improved theories for the mean interaction force.

As for the connection of the thermally activated transport in the
BASEP to the hopping motion in lattice models, one can think of
developing extended jump models. The barrier reduction effect, absent
in the standard ASEP, can, for example, be incorporated by allowing
for different internal states of the particles that correspond to the
different occupancies of the potential wells. Another more obvious
approach is to discretize the potential energy landscape in
space. However, earlier results indicate that in such models current
reversals, absent in the BASEP, become possible
\cite{Chaudhuri/Dhar:2011, Dierl/etal:2014}. Hence, qualitative
features can be different in continuum and related lattice models. A
deeper understanding of the correspondence of continuum and discrete
models in single-file transport should be sought of in future
investigations.

A welcome feature of the BASEP is that it describes physics of biased
single-file motion generated by laser or magnetic fields in confined
geometries, and by flow fields in microfluidic devices. Our results
shall help to interpret experimental findings in such
systems. Experimental studies in open systems offer the possibility to
investigate nonequilibrium phase transitions predicted by theory under
well-controlled conditions. A further interesting aspect, both for
experimental and theoretical work, is to study the implications of
local disturbances in the periodic structure similar as they were
investigated in corresponding lattice models \cite{Kolomeisky:1998,
  Brankov/etal:2004}.

\vspace{2ex}
\acknowledgments
\vspace{-2ex}
Financial support by the Czech Science Foundation (Project
No.\ 17-06716S) and the Deutsche Forschungsgemeinschaft (Project
No.\ 397157593) is gratefully acknowledged.  A.R.\ acknowledges
financial support from the Portuguese Foundation for Science and
Technology (FCT) under Contracts Nos.\ PTDC/FIS-MAC/28146/2017
(LISBOA-01-0145-FEDER-028146) and UID/FIS/00618/2019.  We sincerely
thank the members of the DFG Research Unit FOR 2692 for fruitful
discussions.


\begin{thebibliography}{83}%
\makeatletter
\providecommand \@ifxundefined [1]{%
 \@ifx{#1\undefined}
}%
\providecommand \@ifnum [1]{%
 \ifnum #1\expandafter \@firstoftwo
 \else \expandafter \@secondoftwo
 \fi
}%
\providecommand \@ifx [1]{%
 \ifx #1\expandafter \@firstoftwo
 \else \expandafter \@secondoftwo
 \fi
}%
\providecommand \natexlab [1]{#1}%
\providecommand \enquote  [1]{``#1''}%
\providecommand \bibnamefont  [1]{#1}%
\providecommand \bibfnamefont [1]{#1}%
\providecommand \citenamefont [1]{#1}%
\providecommand \href@noop [0]{\@secondoftwo}%
\providecommand \href [0]{\begingroup \@sanitize@url \@href}%
\providecommand \@href[1]{\@@startlink{#1}\@@href}%
\providecommand \@@href[1]{\endgroup#1\@@endlink}%
\providecommand \@sanitize@url [0]{\catcode `\\12\catcode `\$12\catcode
  `\&12\catcode `\#12\catcode `\^12\catcode `\_12\catcode `\%12\relax}%
\providecommand \@@startlink[1]{}%
\providecommand \@@endlink[0]{}%
\providecommand \url  [0]{\begingroup\@sanitize@url \@url }%
\providecommand \@url [1]{\endgroup\@href {#1}{\urlprefix }}%
\providecommand \urlprefix  [0]{URL }%
\providecommand \Eprint [0]{\href }%
\providecommand \doibase [0]{http://dx.doi.org/}%
\providecommand \selectlanguage [0]{\@gobble}%
\providecommand \bibinfo  [0]{\@secondoftwo}%
\providecommand \bibfield  [0]{\@secondoftwo}%
\providecommand \translation [1]{[#1]}%
\providecommand \BibitemOpen [0]{}%
\providecommand \bibitemStop [0]{}%
\providecommand \bibitemNoStop [0]{.\EOS\space}%
\providecommand \EOS [0]{\spacefactor3000\relax}%
\providecommand \BibitemShut  [1]{\csname bibitem#1\endcsname}%
\let\auto@bib@innerbib\@empty
\bibitem [{\citenamefont {Lipowsky}\ \emph {et~al.}(2001)\citenamefont
  {Lipowsky}, \citenamefont {Klumpp},\ and\ \citenamefont
  {Nieuwenhuizen}}]{Lipowsky/etal:2001}%
  \BibitemOpen
  \bibfield  {author} {\bibinfo {author} {\bibfnamefont {R.}~\bibnamefont
  {Lipowsky}}, \bibinfo {author} {\bibfnamefont {S.}~\bibnamefont {Klumpp}}, \
  and\ \bibinfo {author} {\bibfnamefont {T.~M.}\ \bibnamefont
  {Nieuwenhuizen}},\ }\href@noop {} {\bibfield  {journal} {\bibinfo  {journal}
  {Phys. Rev. Lett.}\ }\textbf {\bibinfo {volume} {87}},\ \bibinfo {pages}
  {108101} (\bibinfo {year} {2001})}\BibitemShut {NoStop}%
\bibitem [{\citenamefont {Frey}\ and\ \citenamefont
  {Kroy}(2005)}]{Frey/Kroy:2005}%
  \BibitemOpen
  \bibfield  {author} {\bibinfo {author} {\bibfnamefont {E.}~\bibnamefont
  {Frey}}\ and\ \bibinfo {author} {\bibfnamefont {K.}~\bibnamefont {Kroy}},\
  }\href@noop {} {\bibfield  {journal} {\bibinfo  {journal} {Ann. Phys.}\
  }\textbf {\bibinfo {volume} {14}},\ \bibinfo {pages} {20} (\bibinfo {year}
  {2005})}\BibitemShut {NoStop}%
\bibitem [{\citenamefont {Hille}(2001)}]{Hille:2001}%
  \BibitemOpen
  \bibfield  {author} {\bibinfo {author} {\bibfnamefont {B.}~\bibnamefont
  {Hille}},\ }\href@noop {} {\emph {\bibinfo {title} {Ionic Channels of
  Excitable Membranes}}},\ \bibinfo {edition} {3rd}\ ed.\ (\bibinfo
  {publisher} {Sinauer Associates, Sunderland},\ \bibinfo {year}
  {2001})\BibitemShut {NoStop}%
\bibitem [{\citenamefont {MacDonald}\ \emph {et~al.}(1968)\citenamefont
  {MacDonald}, \citenamefont {Gibbs},\ and\ \citenamefont
  {Pipkin}}]{MacDonald/etal:1968}%
  \BibitemOpen
  \bibfield  {author} {\bibinfo {author} {\bibfnamefont {C.~T.}\ \bibnamefont
  {MacDonald}}, \bibinfo {author} {\bibfnamefont {J.~H.}\ \bibnamefont
  {Gibbs}}, \ and\ \bibinfo {author} {\bibfnamefont {A.~C.}\ \bibnamefont
  {Pipkin}},\ }\href@noop {} {\bibfield  {journal} {\bibinfo  {journal}
  {Biopolymers}\ }\textbf {\bibinfo {volume} {6}},\ \bibinfo {pages} {1}
  (\bibinfo {year} {1968})}\BibitemShut {NoStop}%
\bibitem [{\citenamefont {Zeng}\ \emph {et~al.}(2018)\citenamefont {Zeng},
  \citenamefont {Chen}, \citenamefont {Wang}, \citenamefont {Zhou},
  \citenamefont {Chen},\ and\ \citenamefont {Dai}}]{Zeng/etal:2018}%
  \BibitemOpen
  \bibfield  {author} {\bibinfo {author} {\bibfnamefont {S.}~\bibnamefont
  {Zeng}}, \bibinfo {author} {\bibfnamefont {J.}~\bibnamefont {Chen}}, \bibinfo
  {author} {\bibfnamefont {X.}~\bibnamefont {Wang}}, \bibinfo {author}
  {\bibfnamefont {G.}~\bibnamefont {Zhou}}, \bibinfo {author} {\bibfnamefont
  {L.}~\bibnamefont {Chen}}, \ and\ \bibinfo {author} {\bibfnamefont
  {C.}~\bibnamefont {Dai}},\ }\href@noop {} {\bibfield  {journal} {\bibinfo
  {journal} {J. Phys. Chem. C}\ }\textbf {\bibinfo {volume} {122}},\ \bibinfo
  {pages} {27681} (\bibinfo {year} {2018})}\BibitemShut {NoStop}%
\bibitem [{\citenamefont {{Van de Voorde}}\ and\ \citenamefont
  {Sels}(2017)}]{VanDeVoorde/Sels:2017}%
  \BibitemOpen
  \bibinfo {editor} {\bibfnamefont {M.}~\bibnamefont {{Van de Voorde}}}\ and\
  \bibinfo {editor} {\bibfnamefont {B.}~\bibnamefont {Sels}},\ eds.,\
  \href@noop {} {\emph {\bibinfo {title} {Nanotechnology in Catalysis:
  Applications in the Chemical Industry, Energy Development, and Environment
  Protection}}}\ (\bibinfo  {publisher} {Wiley-VCH, Weinheim},\ \bibinfo {year}
  {2017})\BibitemShut {NoStop}%
\bibitem [{\citenamefont {Hartmann}(2005)}]{Hartmann:2005}%
  \BibitemOpen
  \bibfield  {author} {\bibinfo {author} {\bibfnamefont {M.}~\bibnamefont
  {Hartmann}},\ }\href@noop {} {\bibfield  {journal} {\bibinfo  {journal}
  {Chem. Mater.}\ }\textbf {\bibinfo {volume} {17}},\ \bibinfo {pages} {4577}
  (\bibinfo {year} {2005})}\BibitemShut {NoStop}%
\bibitem [{\citenamefont {Yiu}\ \emph {et~al.}(2001)\citenamefont {Yiu},
  \citenamefont {Botting}, \citenamefont {Botting},\ and\ \citenamefont
  {Wright}}]{Humphrey/etal:2001}%
  \BibitemOpen
  \bibfield  {author} {\bibinfo {author} {\bibfnamefont {H.~H.~P.}\
  \bibnamefont {Yiu}}, \bibinfo {author} {\bibfnamefont {C.~H.}\ \bibnamefont
  {Botting}}, \bibinfo {author} {\bibfnamefont {N.~P.}\ \bibnamefont
  {Botting}}, \ and\ \bibinfo {author} {\bibfnamefont {P.~A.}\ \bibnamefont
  {Wright}},\ }\href@noop {} {\bibfield  {journal} {\bibinfo  {journal} {Phys.
  Chem. Chem. Phys.}\ }\textbf {\bibinfo {volume} {3}},\ \bibinfo {pages}
  {2983} (\bibinfo {year} {2001})}\BibitemShut {NoStop}%
\bibitem [{\citenamefont {Ma}\ \emph {et~al.}(2015)\citenamefont {Ma},
  \citenamefont {Grey}, \citenamefont {Shen}, \citenamefont {Urbakh},
  \citenamefont {Wu}, \citenamefont {Liu}, \citenamefont {Liu},\ and\
  \citenamefont {Zheng}}]{Ma/etal:2015}%
  \BibitemOpen
  \bibfield  {author} {\bibinfo {author} {\bibfnamefont {M.}~\bibnamefont
  {Ma}}, \bibinfo {author} {\bibfnamefont {F.}~\bibnamefont {Grey}}, \bibinfo
  {author} {\bibfnamefont {L.}~\bibnamefont {Shen}}, \bibinfo {author}
  {\bibfnamefont {M.}~\bibnamefont {Urbakh}}, \bibinfo {author} {\bibfnamefont
  {S.}~\bibnamefont {Wu}}, \bibinfo {author} {\bibfnamefont {J.~Z.}\
  \bibnamefont {Liu}}, \bibinfo {author} {\bibfnamefont {Y.}~\bibnamefont
  {Liu}}, \ and\ \bibinfo {author} {\bibfnamefont {Q.}~\bibnamefont {Zheng}},\
  }\href@noop {} {\bibfield  {journal} {\bibinfo  {journal} {Nat.
  Nanotechnol.}\ }\textbf {\bibinfo {volume} {10}},\ \bibinfo {pages} {692}
  (\bibinfo {year} {2015})}\BibitemShut {NoStop}%
\bibitem [{\citenamefont {Taloni}\ \emph {et~al.}(2017)\citenamefont {Taloni},
  \citenamefont {Flomenbom}, \citenamefont {Casta{\~n}eda-Priego},\ and\
  \citenamefont {Marchesoni}}]{Taloni/etal:2017}%
  \BibitemOpen
  \bibfield  {author} {\bibinfo {author} {\bibfnamefont {A.}~\bibnamefont
  {Taloni}}, \bibinfo {author} {\bibfnamefont {O.}~\bibnamefont {Flomenbom}},
  \bibinfo {author} {\bibfnamefont {R.}~\bibnamefont {Casta{\~n}eda-Priego}}, \
  and\ \bibinfo {author} {\bibfnamefont {F.}~\bibnamefont {Marchesoni}},\
  }\href@noop {} {\bibfield  {journal} {\bibinfo  {journal} {Soft Matter}\
  }\textbf {\bibinfo {volume} {13}},\ \bibinfo {pages} {1096} (\bibinfo {year}
  {2017})}\BibitemShut {NoStop}%
\bibitem [{\citenamefont {Harris}(1965)}]{Harris:1965}%
  \BibitemOpen
  \bibfield  {author} {\bibinfo {author} {\bibfnamefont {T.~E.}\ \bibnamefont
  {Harris}},\ }\href@noop {} {\bibfield  {journal} {\bibinfo  {journal} {J.
  Appl. Probab.}\ }\textbf {\bibinfo {volume} {2}},\ \bibinfo {pages} {323}
  (\bibinfo {year} {1965})}\BibitemShut {NoStop}%
\bibitem [{\citenamefont {Hahn}\ \emph {et~al.}(1996)\citenamefont {Hahn},
  \citenamefont {K{\"a}rger},\ and\ \citenamefont {Kukla}}]{Hahn/etal:1996}%
  \BibitemOpen
  \bibfield  {author} {\bibinfo {author} {\bibfnamefont {K.}~\bibnamefont
  {Hahn}}, \bibinfo {author} {\bibfnamefont {J.}~\bibnamefont {K{\"a}rger}}, \
  and\ \bibinfo {author} {\bibfnamefont {V.}~\bibnamefont {Kukla}},\
  }\href@noop {} {\bibfield  {journal} {\bibinfo  {journal} {Phys. Rev. Lett.}\
  }\textbf {\bibinfo {volume} {76}},\ \bibinfo {pages} {2762} (\bibinfo {year}
  {1996})}\BibitemShut {NoStop}%
\bibitem [{\citenamefont {Chmelik}\ \emph {et~al.}(2018)\citenamefont
  {Chmelik}, \citenamefont {Caro}, \citenamefont {Freude}, \citenamefont
  {Haase}, \citenamefont {Valiullin},\ and\ \citenamefont
  {K{\"a}rger}}]{Chmelik/etal:2018}%
  \BibitemOpen
  \bibfield  {author} {\bibinfo {author} {\bibfnamefont {C.}~\bibnamefont
  {Chmelik}}, \bibinfo {author} {\bibfnamefont {J.}~\bibnamefont {Caro}},
  \bibinfo {author} {\bibfnamefont {D.}~\bibnamefont {Freude}}, \bibinfo
  {author} {\bibfnamefont {J.}~\bibnamefont {Haase}}, \bibinfo {author}
  {\bibfnamefont {R.}~\bibnamefont {Valiullin}}, \ and\ \bibinfo {author}
  {\bibfnamefont {J.}~\bibnamefont {K{\"a}rger}},\ }\enquote {\bibinfo {title}
  {Diffusive {S}preading of {M}olecules in {N}anoporous {M}aterials},}\ \
  (\bibinfo  {publisher} {Springer International Publishing},\ \bibinfo
  {address} {Cham},\ \bibinfo {year} {2018})\ Chap.~\bibinfo {chapter} {10},
  pp.\ \bibinfo {pages} {171--202}\BibitemShut {NoStop}%
\bibitem [{\citenamefont {Cheng}\ and\ \citenamefont
  {Bowers}(2007)}]{Cheng/Bowers:2007}%
  \BibitemOpen
  \bibfield  {author} {\bibinfo {author} {\bibfnamefont {C.}~\bibnamefont
  {Cheng}}\ and\ \bibinfo {author} {\bibfnamefont {C.~R.}\ \bibnamefont
  {Bowers}},\ }\href@noop {} {\bibfield  {journal} {\bibinfo  {journal}
  {ChemPhysChem}\ }\textbf {\bibinfo {volume} {8}},\ \bibinfo {pages} {2077}
  (\bibinfo {year} {2007})}\BibitemShut {NoStop}%
\bibitem [{\citenamefont {Dvoyashkin}\ \emph {et~al.}(2014)\citenamefont
  {Dvoyashkin}, \citenamefont {Bhase}, \citenamefont {Mirnazari}, \citenamefont
  {Vasenkov},\ and\ \citenamefont {Bowers}}]{Dvoyashkin/etal:2014}%
  \BibitemOpen
  \bibfield  {author} {\bibinfo {author} {\bibfnamefont {M.}~\bibnamefont
  {Dvoyashkin}}, \bibinfo {author} {\bibfnamefont {H.}~\bibnamefont {Bhase}},
  \bibinfo {author} {\bibfnamefont {N.}~\bibnamefont {Mirnazari}}, \bibinfo
  {author} {\bibfnamefont {S.}~\bibnamefont {Vasenkov}}, \ and\ \bibinfo
  {author} {\bibfnamefont {C.~R.}\ \bibnamefont {Bowers}},\ }\href@noop {}
  {\bibfield  {journal} {\bibinfo  {journal} {Anal. Chem.}\ }\textbf {\bibinfo
  {volume} {86}},\ \bibinfo {pages} {2200} (\bibinfo {year}
  {2014})}\BibitemShut {NoStop}%
\bibitem [{\citenamefont {Wei}\ \emph {et~al.}(2000)\citenamefont {Wei},
  \citenamefont {Bechinger},\ and\ \citenamefont {Leiderer}}]{Wei/etal:2000}%
  \BibitemOpen
  \bibfield  {author} {\bibinfo {author} {\bibfnamefont {Q.-H.}\ \bibnamefont
  {Wei}}, \bibinfo {author} {\bibfnamefont {C.}~\bibnamefont {Bechinger}}, \
  and\ \bibinfo {author} {\bibfnamefont {P.}~\bibnamefont {Leiderer}},\
  }\href@noop {} {\bibfield  {journal} {\bibinfo  {journal} {Science}\ }\textbf
  {\bibinfo {volume} {287}},\ \bibinfo {pages} {625} (\bibinfo {year}
  {2000})}\BibitemShut {NoStop}%
\bibitem [{\citenamefont {Taloni}\ \emph {et~al.}(2010)\citenamefont {Taloni},
  \citenamefont {Chechkin},\ and\ \citenamefont {Klafter}}]{Taloni/etal:2010}%
  \BibitemOpen
  \bibfield  {author} {\bibinfo {author} {\bibfnamefont {A.}~\bibnamefont
  {Taloni}}, \bibinfo {author} {\bibfnamefont {A.}~\bibnamefont {Chechkin}}, \
  and\ \bibinfo {author} {\bibfnamefont {J.}~\bibnamefont {Klafter}},\
  }\href@noop {} {\bibfield  {journal} {\bibinfo  {journal} {Phys. Rev. Lett.}\
  }\textbf {\bibinfo {volume} {104}},\ \bibinfo {pages} {160602} (\bibinfo
  {year} {2010})}\BibitemShut {NoStop}%
\bibitem [{\citenamefont {Sanders}\ and\ \citenamefont
  {Ambj{\"o}rnsson}(2012)}]{Sanders/Ambjornsson:2012}%
  \BibitemOpen
  \bibfield  {author} {\bibinfo {author} {\bibfnamefont {L.~P.}\ \bibnamefont
  {Sanders}}\ and\ \bibinfo {author} {\bibfnamefont {T.}~\bibnamefont
  {Ambj{\"o}rnsson}},\ }\href@noop {} {\bibfield  {journal} {\bibinfo
  {journal} {J. Chem. Phys.}\ }\textbf {\bibinfo {volume} {136}},\ \bibinfo
  {pages} {175103} (\bibinfo {year} {2012})}\BibitemShut {NoStop}%
\bibitem [{\citenamefont {Leibovich}\ and\ \citenamefont
  {Barkai}(2013)}]{Leibovich/Barkai:2013}%
  \BibitemOpen
  \bibfield  {author} {\bibinfo {author} {\bibfnamefont {N.}~\bibnamefont
  {Leibovich}}\ and\ \bibinfo {author} {\bibfnamefont {E.}~\bibnamefont
  {Barkai}},\ }\href@noop {} {\bibfield  {journal} {\bibinfo  {journal} {Phys.
  Rev. E}\ }\textbf {\bibinfo {volume} {88}},\ \bibinfo {pages} {032107}
  (\bibinfo {year} {2013})}\BibitemShut {NoStop}%
\bibitem [{\citenamefont {Lizana}\ and\ \citenamefont
  {Ambj\"ornsson}(2008)}]{Lizana/Ambjornsson:2008}%
  \BibitemOpen
  \bibfield  {author} {\bibinfo {author} {\bibfnamefont {L.}~\bibnamefont
  {Lizana}}\ and\ \bibinfo {author} {\bibfnamefont {T.}~\bibnamefont
  {Ambj\"ornsson}},\ }\href@noop {} {\bibfield  {journal} {\bibinfo  {journal}
  {Phys. Rev. Lett.}\ }\textbf {\bibinfo {volume} {100}},\ \bibinfo {pages}
  {200601} (\bibinfo {year} {2008})}\BibitemShut {NoStop}%
\bibitem [{\citenamefont {Barkai}\ and\ \citenamefont
  {Silbey}(2009)}]{Barkai/Silbey:2009}%
  \BibitemOpen
  \bibfield  {author} {\bibinfo {author} {\bibfnamefont {E.}~\bibnamefont
  {Barkai}}\ and\ \bibinfo {author} {\bibfnamefont {R.}~\bibnamefont
  {Silbey}},\ }\href@noop {} {\bibfield  {journal} {\bibinfo  {journal} {Phys.
  Rev. Lett.}\ }\textbf {\bibinfo {volume} {102}},\ \bibinfo {pages} {050602}
  (\bibinfo {year} {2009})}\BibitemShut {NoStop}%
\bibitem [{\citenamefont {Ryabov}\ and\ \citenamefont
  {Chvosta}(2011)}]{Ryabov/Chvosta:2011}%
  \BibitemOpen
  \bibfield  {author} {\bibinfo {author} {\bibfnamefont {A.}~\bibnamefont
  {Ryabov}}\ and\ \bibinfo {author} {\bibfnamefont {P.}~\bibnamefont
  {Chvosta}},\ }\href@noop {} {\bibfield  {journal} {\bibinfo  {journal} {Phys.
  Rev. E}\ }\textbf {\bibinfo {volume} {83}},\ \bibinfo {pages} {020106(R)}
  (\bibinfo {year} {2011})}\BibitemShut {NoStop}%
\bibitem [{\citenamefont {Ryabov}\ and\ \citenamefont
  {Chvosta}(2012)}]{Ryabov/Chvosta:2012}%
  \BibitemOpen
  \bibfield  {author} {\bibinfo {author} {\bibfnamefont {A.}~\bibnamefont
  {Ryabov}}\ and\ \bibinfo {author} {\bibfnamefont {P.}~\bibnamefont
  {Chvosta}},\ }\href@noop {} {\bibfield  {journal} {\bibinfo  {journal} {J.
  Chem. Phys.}\ }\textbf {\bibinfo {volume} {136}},\ \bibinfo {pages} {064114}
  (\bibinfo {year} {2012})}\BibitemShut {NoStop}%
\bibitem [{\citenamefont {Ryabov}(2013)}]{Ryabov:2013}%
  \BibitemOpen
  \bibfield  {author} {\bibinfo {author} {\bibfnamefont {A.}~\bibnamefont
  {Ryabov}},\ }\href@noop {} {\bibfield  {journal} {\bibinfo  {journal} {J.
  Chem. Phys.}\ }\textbf {\bibinfo {volume} {138}},\ \bibinfo {pages} {154104}
  (\bibinfo {year} {2013})}\BibitemShut {NoStop}%
\bibitem [{\citenamefont {Locatelli}\ \emph {et~al.}(2016)\citenamefont
  {Locatelli}, \citenamefont {Pierno}, \citenamefont {Baldovin}, \citenamefont
  {Orlandini}, \citenamefont {Tan},\ and\ \citenamefont
  {Pagliara}}]{Locatelli/etal:2016}%
  \BibitemOpen
  \bibfield  {author} {\bibinfo {author} {\bibfnamefont {E.}~\bibnamefont
  {Locatelli}}, \bibinfo {author} {\bibfnamefont {M.}~\bibnamefont {Pierno}},
  \bibinfo {author} {\bibfnamefont {F.}~\bibnamefont {Baldovin}}, \bibinfo
  {author} {\bibfnamefont {E.}~\bibnamefont {Orlandini}}, \bibinfo {author}
  {\bibfnamefont {Y.}~\bibnamefont {Tan}}, \ and\ \bibinfo {author}
  {\bibfnamefont {S.}~\bibnamefont {Pagliara}},\ }\href@noop {} {\bibfield
  {journal} {\bibinfo  {journal} {Phys. Rev. Lett.}\ }\textbf {\bibinfo
  {volume} {117}},\ \bibinfo {pages} {038001} (\bibinfo {year}
  {2016})}\BibitemShut {NoStop}%
\bibitem [{\citenamefont {Krapivsky}\ \emph {et~al.}(2015)\citenamefont
  {Krapivsky}, \citenamefont {Mallick},\ and\ \citenamefont
  {Sadhu}}]{Krapivsky/etal:2015}%
  \BibitemOpen
  \bibfield  {author} {\bibinfo {author} {\bibfnamefont {P.~L.}\ \bibnamefont
  {Krapivsky}}, \bibinfo {author} {\bibfnamefont {K.}~\bibnamefont {Mallick}},
  \ and\ \bibinfo {author} {\bibfnamefont {T.}~\bibnamefont {Sadhu}},\
  }\href@noop {} {\bibfield  {journal} {\bibinfo  {journal} {J. Stat. Mech.}\
  }\textbf {\bibinfo {volume} {2015}},\ \bibinfo {pages} {P09007} (\bibinfo
  {year} {2015})}\BibitemShut {NoStop}%
\bibitem [{\citenamefont {Mon}\ and\ \citenamefont
  {Percus}(2002)}]{Mon/Percus:2002}%
  \BibitemOpen
  \bibfield  {author} {\bibinfo {author} {\bibfnamefont {K.~K.}\ \bibnamefont
  {Mon}}\ and\ \bibinfo {author} {\bibfnamefont {J.~K.}\ \bibnamefont
  {Percus}},\ }\href@noop {} {\bibfield  {journal} {\bibinfo  {journal} {J.
  Chem. Phys.}\ }\textbf {\bibinfo {volume} {117}},\ \bibinfo {pages} {2289}
  (\bibinfo {year} {2002})}\BibitemShut {NoStop}%
\bibitem [{\citenamefont {Ooshida}\ \emph {et~al.}(2018)\citenamefont
  {Ooshida}, \citenamefont {Goto},\ and\ \citenamefont
  {Otsuki}}]{Ooshida/etal:2018}%
  \BibitemOpen
  \bibfield  {author} {\bibinfo {author} {\bibfnamefont {T.}~\bibnamefont
  {Ooshida}}, \bibinfo {author} {\bibfnamefont {S.}~\bibnamefont {Goto}}, \
  and\ \bibinfo {author} {\bibfnamefont {M.}~\bibnamefont {Otsuki}},\
  }\href@noop {} {\bibfield  {journal} {\bibinfo  {journal} {Entropy}\ }\textbf
  {\bibinfo {volume} {20}},\ \bibinfo {pages} {565} (\bibinfo {year}
  {2018})}\BibitemShut {NoStop}%
\bibitem [{\citenamefont {Ahmadi}\ \emph {et~al.}(2019)\citenamefont {Ahmadi},
  \citenamefont {Schmidt}, \citenamefont {Spiteri},\ and\ \citenamefont
  {Bowles}}]{Ahmadi/etal:2019}%
  \BibitemOpen
  \bibfield  {author} {\bibinfo {author} {\bibfnamefont {S.}~\bibnamefont
  {Ahmadi}}, \bibinfo {author} {\bibfnamefont {M.}~\bibnamefont {Schmidt}},
  \bibinfo {author} {\bibfnamefont {R.~J.}\ \bibnamefont {Spiteri}}, \ and\
  \bibinfo {author} {\bibfnamefont {R.~K.}\ \bibnamefont {Bowles}},\
  }\href@noop {} {\bibfield  {journal} {\bibinfo  {journal} {J. Chem. Phys.}\
  }\textbf {\bibinfo {volume} {150}},\ \bibinfo {pages} {224501} (\bibinfo
  {year} {2019})}\BibitemShut {NoStop}%
\bibitem [{\citenamefont {Krapivsky}\ \emph {et~al.}(2014)\citenamefont
  {Krapivsky}, \citenamefont {Mallick},\ and\ \citenamefont
  {Sadhu}}]{Krapivsky/etal:2014}%
  \BibitemOpen
  \bibfield  {author} {\bibinfo {author} {\bibfnamefont {P.~L.}\ \bibnamefont
  {Krapivsky}}, \bibinfo {author} {\bibfnamefont {K.}~\bibnamefont {Mallick}},
  \ and\ \bibinfo {author} {\bibfnamefont {T.}~\bibnamefont {Sadhu}},\
  }\href@noop {} {\bibfield  {journal} {\bibinfo  {journal} {Phys. Rev. Lett.}\
  }\textbf {\bibinfo {volume} {113}},\ \bibinfo {pages} {078101} (\bibinfo
  {year} {2014})}\BibitemShut {NoStop}%
\bibitem [{\citenamefont {Schadschneider}\ \emph {et~al.}(2010)\citenamefont
  {Schadschneider}, \citenamefont {Chowdhury},\ and\ \citenamefont
  {Nishinari}}]{Schadschneider/etal:2010}%
  \BibitemOpen
  \bibfield  {author} {\bibinfo {author} {\bibfnamefont {A.}~\bibnamefont
  {Schadschneider}}, \bibinfo {author} {\bibfnamefont {D.}~\bibnamefont
  {Chowdhury}}, \ and\ \bibinfo {author} {\bibfnamefont {K.}~\bibnamefont
  {Nishinari}},\ }\href@noop {} {\emph {\bibinfo {title} {Stochastic Transport
  in Complex Systems: From Molecules to Vehicles}}},\ \bibinfo {edition} {3rd}\
  ed.\ (\bibinfo  {publisher} {Elsevier Science, Amsterdam},\ \bibinfo {year}
  {2010})\BibitemShut {NoStop}%
\bibitem [{\citenamefont {Chou}\ \emph {et~al.}(2011)\citenamefont {Chou},
  \citenamefont {Mallick},\ and\ \citenamefont {Zia}}]{Chou/etal:2011}%
  \BibitemOpen
  \bibfield  {author} {\bibinfo {author} {\bibfnamefont {T.}~\bibnamefont
  {Chou}}, \bibinfo {author} {\bibfnamefont {K.}~\bibnamefont {Mallick}}, \
  and\ \bibinfo {author} {\bibfnamefont {R.~K.~P.}\ \bibnamefont {Zia}},\
  }\href@noop {} {\bibfield  {journal} {\bibinfo  {journal} {Rep. Prog. Phys.}\
  }\textbf {\bibinfo {volume} {74}},\ \bibinfo {pages} {116601} (\bibinfo
  {year} {2011})}\BibitemShut {NoStop}%
\bibitem [{\citenamefont {Kolomeisky}(2013)}]{Kolomeisky:2013}%
  \BibitemOpen
  \bibfield  {author} {\bibinfo {author} {\bibfnamefont {A.~B.}\ \bibnamefont
  {Kolomeisky}},\ }\href@noop {} {\bibfield  {journal} {\bibinfo  {journal} {J.
  Phys.: Condens. Matter}\ }\textbf {\bibinfo {volume} {25}},\ \bibinfo {pages}
  {463101} (\bibinfo {year} {2013})}\BibitemShut {NoStop}%
\bibitem [{\citenamefont {Appert-Rolland}\ \emph {et~al.}(2015)\citenamefont
  {Appert-Rolland}, \citenamefont {Ebbinghaus},\ and\ \citenamefont
  {Santen}}]{Appert-Rolland/etal:2015}%
  \BibitemOpen
  \bibfield  {author} {\bibinfo {author} {\bibfnamefont {C.}~\bibnamefont
  {Appert-Rolland}}, \bibinfo {author} {\bibfnamefont {M.}~\bibnamefont
  {Ebbinghaus}}, \ and\ \bibinfo {author} {\bibfnamefont {L.}~\bibnamefont
  {Santen}},\ }\href@noop {} {\bibfield  {journal} {\bibinfo  {journal} {Phys.
  Rep.}\ }\textbf {\bibinfo {volume} {593}},\ \bibinfo {pages} {1 } (\bibinfo
  {year} {2015})}\BibitemShut {NoStop}%
\bibitem [{\citenamefont {Derrida}(1998)}]{Derrida:1998}%
  \BibitemOpen
  \bibfield  {author} {\bibinfo {author} {\bibfnamefont {B.}~\bibnamefont
  {Derrida}},\ }\href@noop {} {\bibfield  {journal} {\bibinfo  {journal} {Phys.
  Rep.}\ }\textbf {\bibinfo {volume} {301}},\ \bibinfo {pages} {65} (\bibinfo
  {year} {1998})}\BibitemShut {NoStop}%
\bibitem [{\citenamefont {Sch{\"u}tz}(2001)}]{Schuetz:2001}%
  \BibitemOpen
  \bibfield  {author} {\bibinfo {author} {\bibfnamefont {G.~M.}\ \bibnamefont
  {Sch{\"u}tz}},\ }in\ \href@noop {} {\emph {\bibinfo {booktitle} {Phase
  Transitions and Critical Phenomena}}},\ Vol.~\bibinfo {volume} {19},\
  \bibinfo {editor} {edited by\ \bibinfo {editor} {\bibfnamefont
  {C.}~\bibnamefont {Domb}}\ and\ \bibinfo {editor} {\bibfnamefont
  {J.}~\bibnamefont {Lebowitz}}}\ (\bibinfo  {publisher} {Academic Press},\
  \bibinfo {address} {London},\ \bibinfo {year} {2001})\ pp.\ \bibinfo {pages}
  {1--251}\BibitemShut {NoStop}%
\bibitem [{\citenamefont {Blythe}\ and\ \citenamefont
  {Evans}(2007)}]{Blythe/Evans:2007}%
  \BibitemOpen
  \bibfield  {author} {\bibinfo {author} {\bibfnamefont {R.~A.}\ \bibnamefont
  {Blythe}}\ and\ \bibinfo {author} {\bibfnamefont {M.~R.}\ \bibnamefont
  {Evans}},\ }\href@noop {} {\bibfield  {journal} {\bibinfo  {journal} {J.
  Phys. A: Math. Theor.}\ }\textbf {\bibinfo {volume} {40}},\ \bibinfo {pages}
  {R333} (\bibinfo {year} {2007})}\BibitemShut {NoStop}%
\bibitem [{\citenamefont {Krug}(1991)}]{Krug:1991}%
  \BibitemOpen
  \bibfield  {author} {\bibinfo {author} {\bibfnamefont {J.}~\bibnamefont
  {Krug}},\ }\href@noop {} {\bibfield  {journal} {\bibinfo  {journal} {Phys.
  Rev. Lett.}\ }\textbf {\bibinfo {volume} {67}},\ \bibinfo {pages} {1882}
  (\bibinfo {year} {1991})}\BibitemShut {NoStop}%
\bibitem [{\citenamefont {Parmeggiani}\ \emph {et~al.}(2003)\citenamefont
  {Parmeggiani}, \citenamefont {Franosch},\ and\ \citenamefont
  {Frey}}]{Parmeggiani/etal:2003}%
  \BibitemOpen
  \bibfield  {author} {\bibinfo {author} {\bibfnamefont {A.}~\bibnamefont
  {Parmeggiani}}, \bibinfo {author} {\bibfnamefont {T.}~\bibnamefont
  {Franosch}}, \ and\ \bibinfo {author} {\bibfnamefont {E.}~\bibnamefont
  {Frey}},\ }\href@noop {} {\bibfield  {journal} {\bibinfo  {journal} {Phys.
  Rev. Lett.}\ }\textbf {\bibinfo {volume} {90}},\ \bibinfo {pages} {086601}
  (\bibinfo {year} {2003})}\BibitemShut {NoStop}%
\bibitem [{\citenamefont {Antal}\ and\ \citenamefont
  {Sch\"utz}(2000)}]{Antal/Schuetz:2000}%
  \BibitemOpen
  \bibfield  {author} {\bibinfo {author} {\bibfnamefont {T.}~\bibnamefont
  {Antal}}\ and\ \bibinfo {author} {\bibfnamefont {G.~M.}\ \bibnamefont
  {Sch\"utz}},\ }\href@noop {} {\bibfield  {journal} {\bibinfo  {journal}
  {Phys. Rev. E}\ }\textbf {\bibinfo {volume} {62}},\ \bibinfo {pages} {83}
  (\bibinfo {year} {2000})}\BibitemShut {NoStop}%
\bibitem [{\citenamefont {Dierl}\ \emph {et~al.}(2011)\citenamefont {Dierl},
  \citenamefont {Maass},\ and\ \citenamefont {Einax}}]{Dierl/etal:2011}%
  \BibitemOpen
  \bibfield  {author} {\bibinfo {author} {\bibfnamefont {M.}~\bibnamefont
  {Dierl}}, \bibinfo {author} {\bibfnamefont {P.}~\bibnamefont {Maass}}, \ and\
  \bibinfo {author} {\bibfnamefont {M.}~\bibnamefont {Einax}},\ }\href@noop {}
  {\bibfield  {journal} {\bibinfo  {journal} {EPL}\ }\textbf {\bibinfo {volume}
  {93}},\ \bibinfo {pages} {50003} (\bibinfo {year} {2011})}\BibitemShut
  {NoStop}%
\bibitem [{\citenamefont {Dierl}\ \emph {et~al.}(2012)\citenamefont {Dierl},
  \citenamefont {Maass},\ and\ \citenamefont {Einax}}]{Dierl/etal:2012}%
  \BibitemOpen
  \bibfield  {author} {\bibinfo {author} {\bibfnamefont {M.}~\bibnamefont
  {Dierl}}, \bibinfo {author} {\bibfnamefont {P.}~\bibnamefont {Maass}}, \ and\
  \bibinfo {author} {\bibfnamefont {M.}~\bibnamefont {Einax}},\ }\href@noop {}
  {\bibfield  {journal} {\bibinfo  {journal} {Phys. Rev. Lett.}\ }\textbf
  {\bibinfo {volume} {108}},\ \bibinfo {pages} {060603} (\bibinfo {year}
  {2012})}\BibitemShut {NoStop}%
\bibitem [{\citenamefont {Dierl}\ \emph {et~al.}(2013)\citenamefont {Dierl},
  \citenamefont {Einax},\ and\ \citenamefont {Maass}}]{Dierl/etal:2013}%
  \BibitemOpen
  \bibfield  {author} {\bibinfo {author} {\bibfnamefont {M.}~\bibnamefont
  {Dierl}}, \bibinfo {author} {\bibfnamefont {M.}~\bibnamefont {Einax}}, \ and\
  \bibinfo {author} {\bibfnamefont {P.}~\bibnamefont {Maass}},\ }\href@noop {}
  {\bibfield  {journal} {\bibinfo  {journal} {Phys. Rev. E}\ }\textbf {\bibinfo
  {volume} {87}},\ \bibinfo {pages} {062126} (\bibinfo {year}
  {2013})}\BibitemShut {NoStop}%
\bibitem [{\citenamefont {Dierl}\ \emph {et~al.}(2014)\citenamefont {Dierl},
  \citenamefont {Dieterich}, \citenamefont {Einax},\ and\ \citenamefont
  {Maass}}]{Dierl/etal:2014}%
  \BibitemOpen
  \bibfield  {author} {\bibinfo {author} {\bibfnamefont {M.}~\bibnamefont
  {Dierl}}, \bibinfo {author} {\bibfnamefont {W.}~\bibnamefont {Dieterich}},
  \bibinfo {author} {\bibfnamefont {M.}~\bibnamefont {Einax}}, \ and\ \bibinfo
  {author} {\bibfnamefont {P.}~\bibnamefont {Maass}},\ }\href@noop {}
  {\bibfield  {journal} {\bibinfo  {journal} {Phys. Rev. Lett.}\ }\textbf
  {\bibinfo {volume} {112}},\ \bibinfo {pages} {150601} (\bibinfo {year}
  {2014})}\BibitemShut {NoStop}%
\bibitem [{\citenamefont {Bertini}\ \emph {et~al.}(2005)\citenamefont
  {Bertini}, \citenamefont {De~Sole}, \citenamefont {Gabrielli}, \citenamefont
  {Jona-Lasinio},\ and\ \citenamefont {Landim}}]{Bertini/etal:2005}%
  \BibitemOpen
  \bibfield  {author} {\bibinfo {author} {\bibfnamefont {L.}~\bibnamefont
  {Bertini}}, \bibinfo {author} {\bibfnamefont {A.}~\bibnamefont {De~Sole}},
  \bibinfo {author} {\bibfnamefont {D.}~\bibnamefont {Gabrielli}}, \bibinfo
  {author} {\bibfnamefont {G.}~\bibnamefont {Jona-Lasinio}}, \ and\ \bibinfo
  {author} {\bibfnamefont {C.}~\bibnamefont {Landim}},\ }\href@noop {}
  {\bibfield  {journal} {\bibinfo  {journal} {Phys. Rev. Lett.}\ }\textbf
  {\bibinfo {volume} {94}},\ \bibinfo {pages} {030601} (\bibinfo {year}
  {2005})}\BibitemShut {NoStop}%
\bibitem [{\citenamefont {Lazarescu}(2015)}]{Lazarescu:2015}%
  \BibitemOpen
  \bibfield  {author} {\bibinfo {author} {\bibfnamefont {A.}~\bibnamefont
  {Lazarescu}},\ }\href@noop {} {\bibfield  {journal} {\bibinfo  {journal} {J.
  Phys. A: Math. Theor.}\ }\textbf {\bibinfo {volume} {48}},\ \bibinfo {pages}
  {503001} (\bibinfo {year} {2015})}\BibitemShut {NoStop}%
\bibitem [{\citenamefont {Baek}\ \emph {et~al.}(2017)\citenamefont {Baek},
  \citenamefont {Kafri},\ and\ \citenamefont {Lecomte}}]{Baek/etal:2017}%
  \BibitemOpen
  \bibfield  {author} {\bibinfo {author} {\bibfnamefont {Y.}~\bibnamefont
  {Baek}}, \bibinfo {author} {\bibfnamefont {Y.}~\bibnamefont {Kafri}}, \ and\
  \bibinfo {author} {\bibfnamefont {V.}~\bibnamefont {Lecomte}},\ }\href@noop
  {} {\bibfield  {journal} {\bibinfo  {journal} {Phys. Rev. Lett.}\ }\textbf
  {\bibinfo {volume} {118}},\ \bibinfo {pages} {030604} (\bibinfo {year}
  {2017})}\BibitemShut {NoStop}%
\bibitem [{\citenamefont {Evans}(1996)}]{Evans:1996}%
  \BibitemOpen
  \bibfield  {author} {\bibinfo {author} {\bibfnamefont {M.~R.}\ \bibnamefont
  {Evans}},\ }\href@noop {} {\bibfield  {journal} {\bibinfo  {journal} {EPL}\
  }\textbf {\bibinfo {volume} {36}},\ \bibinfo {pages} {13} (\bibinfo {year}
  {1996})}\BibitemShut {NoStop}%
\bibitem [{\citenamefont {Concannon}\ and\ \citenamefont
  {Blythe}(2014)}]{Concannon/Blythe:2014}%
  \BibitemOpen
  \bibfield  {author} {\bibinfo {author} {\bibfnamefont {R.~J.}\ \bibnamefont
  {Concannon}}\ and\ \bibinfo {author} {\bibfnamefont {R.~A.}\ \bibnamefont
  {Blythe}},\ }\href@noop {} {\bibfield  {journal} {\bibinfo  {journal} {Phys.
  Rev. Lett.}\ }\textbf {\bibinfo {volume} {112}},\ \bibinfo {pages} {050603}
  (\bibinfo {year} {2014})}\BibitemShut {NoStop}%
\bibitem [{\citenamefont {Popkov}\ \emph {et~al.}(2015)\citenamefont {Popkov},
  \citenamefont {Schadschneider}, \citenamefont {Schmidt},\ and\ \citenamefont
  {Sch\"utz}}]{Popkov/etal:2015}%
  \BibitemOpen
  \bibfield  {author} {\bibinfo {author} {\bibfnamefont {V.}~\bibnamefont
  {Popkov}}, \bibinfo {author} {\bibfnamefont {A.}~\bibnamefont
  {Schadschneider}}, \bibinfo {author} {\bibfnamefont {J.}~\bibnamefont
  {Schmidt}}, \ and\ \bibinfo {author} {\bibfnamefont {G.~M.}\ \bibnamefont
  {Sch\"utz}},\ }\href@noop {} {\bibfield  {journal} {\bibinfo  {journal}
  {PNAS}\ }\textbf {\bibinfo {volume} {112}},\ \bibinfo {pages} {12645}
  (\bibinfo {year} {2015})}\BibitemShut {NoStop}%
\bibitem [{\citenamefont {Chen}\ \emph {et~al.}(2018)\citenamefont {Chen},
  \citenamefont {de~Gier}, \citenamefont {Hiki},\ and\ \citenamefont
  {Sasamoto}}]{Chen/etal:2018}%
  \BibitemOpen
  \bibfield  {author} {\bibinfo {author} {\bibfnamefont {Z.}~\bibnamefont
  {Chen}}, \bibinfo {author} {\bibfnamefont {J.}~\bibnamefont {de~Gier}},
  \bibinfo {author} {\bibfnamefont {I.}~\bibnamefont {Hiki}}, \ and\ \bibinfo
  {author} {\bibfnamefont {T.}~\bibnamefont {Sasamoto}},\ }\href@noop {}
  {\bibfield  {journal} {\bibinfo  {journal} {Phys. Rev. Lett.}\ }\textbf
  {\bibinfo {volume} {120}},\ \bibinfo {pages} {240601} (\bibinfo {year}
  {2018})}\BibitemShut {NoStop}%
\bibitem [{\citenamefont {Lips}\ \emph {et~al.}(2018)\citenamefont {Lips},
  \citenamefont {Ryabov},\ and\ \citenamefont {Maass}}]{Lips/etal:2018}%
  \BibitemOpen
  \bibfield  {author} {\bibinfo {author} {\bibfnamefont {D.}~\bibnamefont
  {Lips}}, \bibinfo {author} {\bibfnamefont {A.}~\bibnamefont {Ryabov}}, \ and\
  \bibinfo {author} {\bibfnamefont {P.}~\bibnamefont {Maass}},\ }\href@noop {}
  {\bibfield  {journal} {\bibinfo  {journal} {Phys. Rev. Lett.}\ }\textbf
  {\bibinfo {volume} {121}},\ \bibinfo {pages} {160601} (\bibinfo {year}
  {2018})}\BibitemShut {NoStop}%
\bibitem [{\citenamefont {Arzola}\ \emph {et~al.}(2017)\citenamefont {Arzola},
  \citenamefont {Villasante-Barahona}, \citenamefont {Volke-Sep{\'u}lveda},
  \citenamefont {J{\'a}kl},\ and\ \citenamefont
  {Zem{\'a}nek}}]{Arzola/etal:2017}%
  \BibitemOpen
  \bibfield  {author} {\bibinfo {author} {\bibfnamefont {A.~V.}\ \bibnamefont
  {Arzola}}, \bibinfo {author} {\bibfnamefont {M.}~\bibnamefont
  {Villasante-Barahona}}, \bibinfo {author} {\bibfnamefont {K.}~\bibnamefont
  {Volke-Sep{\'u}lveda}}, \bibinfo {author} {\bibfnamefont {P.}~\bibnamefont
  {J{\'a}kl}}, \ and\ \bibinfo {author} {\bibfnamefont {P.}~\bibnamefont
  {Zem{\'a}nek}},\ }\href@noop {} {\bibfield  {journal} {\bibinfo  {journal}
  {Phys. Rev. Lett.}\ }\textbf {\bibinfo {volume} {118}},\ \bibinfo {pages}
  {138002} (\bibinfo {year} {2017})}\BibitemShut {NoStop}%
\bibitem [{\citenamefont {Skaug}\ \emph {et~al.}(2018)\citenamefont {Skaug},
  \citenamefont {Schwemmer}, \citenamefont {Fringes}, \citenamefont
  {Rawlings},\ and\ \citenamefont {Knoll}}]{Skaug/etal:2018}%
  \BibitemOpen
  \bibfield  {author} {\bibinfo {author} {\bibfnamefont {M.~J.}\ \bibnamefont
  {Skaug}}, \bibinfo {author} {\bibfnamefont {C.}~\bibnamefont {Schwemmer}},
  \bibinfo {author} {\bibfnamefont {S.}~\bibnamefont {Fringes}}, \bibinfo
  {author} {\bibfnamefont {C.~D.}\ \bibnamefont {Rawlings}}, \ and\ \bibinfo
  {author} {\bibfnamefont {A.~W.}\ \bibnamefont {Knoll}},\ }\href@noop {}
  {\bibfield  {journal} {\bibinfo  {journal} {Science}\ }\textbf {\bibinfo
  {volume} {359}},\ \bibinfo {pages} {1505} (\bibinfo {year}
  {2018})}\BibitemShut {NoStop}%
\bibitem [{\citenamefont {Schwemmer}\ \emph {et~al.}(2018)\citenamefont
  {Schwemmer}, \citenamefont {Fringes}, \citenamefont {Duerig}, \citenamefont
  {Ryu},\ and\ \citenamefont {Knoll}}]{Schwemmer/etal:2018}%
  \BibitemOpen
  \bibfield  {author} {\bibinfo {author} {\bibfnamefont {C.}~\bibnamefont
  {Schwemmer}}, \bibinfo {author} {\bibfnamefont {S.}~\bibnamefont {Fringes}},
  \bibinfo {author} {\bibfnamefont {U.}~\bibnamefont {Duerig}}, \bibinfo
  {author} {\bibfnamefont {Y.~K.}\ \bibnamefont {Ryu}}, \ and\ \bibinfo
  {author} {\bibfnamefont {A.~W.}\ \bibnamefont {Knoll}},\ }\href@noop {}
  {\bibfield  {journal} {\bibinfo  {journal} {Phys. Rev. Lett.}\ }\textbf
  {\bibinfo {volume} {121}},\ \bibinfo {pages} {104102} (\bibinfo {year}
  {2018})}\BibitemShut {NoStop}%
\bibitem [{\citenamefont {Stoop}\ \emph {et~al.}(2019)\citenamefont {Stoop},
  \citenamefont {Straube},\ and\ \citenamefont {Tierno}}]{Stoop/etal:2018}%
  \BibitemOpen
  \bibfield  {author} {\bibinfo {author} {\bibfnamefont {R.}~\bibnamefont
  {Stoop}}, \bibinfo {author} {\bibfnamefont {A.}~\bibnamefont {Straube}}, \
  and\ \bibinfo {author} {\bibfnamefont {P.}~\bibnamefont {Tierno}},\
  }\href@noop {} {\bibfield  {journal} {\bibinfo  {journal} {Nano Lett.}\
  }\textbf {\bibinfo {volume} {19}},\ \bibinfo {pages} {433} (\bibinfo {year}
  {2019})}\BibitemShut {NoStop}%
\bibitem [{\citenamefont {Misiunas}\ and\ \citenamefont
  {Keyser}(2019)}]{Misiunas/Keyser:2019}%
  \BibitemOpen
  \bibfield  {author} {\bibinfo {author} {\bibfnamefont {K.}~\bibnamefont
  {Misiunas}}\ and\ \bibinfo {author} {\bibfnamefont {U.~F.}\ \bibnamefont
  {Keyser}},\ }\href@noop {} {\bibfield  {journal} {\bibinfo  {journal} {Phys.
  Rev. Lett.}\ }\textbf {\bibinfo {volume} {122}},\ \bibinfo {pages} {214501}
  (\bibinfo {year} {2019})}\BibitemShut {NoStop}%
\bibitem [{\citenamefont {Straube}\ and\ \citenamefont
  {Tierno}(2013)}]{Straube/Tierno:2013}%
  \BibitemOpen
  \bibfield  {author} {\bibinfo {author} {\bibfnamefont {A.~V.}\ \bibnamefont
  {Straube}}\ and\ \bibinfo {author} {\bibfnamefont {P.}~\bibnamefont
  {Tierno}},\ }\href@noop {} {\bibfield  {journal} {\bibinfo  {journal} {EPL}\
  }\textbf {\bibinfo {volume} {103}},\ \bibinfo {pages} {28001} (\bibinfo
  {year} {2013})}\BibitemShut {NoStop}%
\bibitem [{\citenamefont {Jain}\ \emph {et~al.}(2007)\citenamefont {Jain},
  \citenamefont {Marathe}, \citenamefont {Chaudhuri},\ and\ \citenamefont
  {Dhar}}]{Jain/etal:2007}%
  \BibitemOpen
  \bibfield  {author} {\bibinfo {author} {\bibfnamefont {K.}~\bibnamefont
  {Jain}}, \bibinfo {author} {\bibfnamefont {R.}~\bibnamefont {Marathe}},
  \bibinfo {author} {\bibfnamefont {A.}~\bibnamefont {Chaudhuri}}, \ and\
  \bibinfo {author} {\bibfnamefont {A.}~\bibnamefont {Dhar}},\ }\href@noop {}
  {\bibfield  {journal} {\bibinfo  {journal} {Phys. Rev. Lett.}\ }\textbf
  {\bibinfo {volume} {99}},\ \bibinfo {pages} {190601} (\bibinfo {year}
  {2007})}\BibitemShut {NoStop}%
\bibitem [{\citenamefont {Slanina}(2009{\natexlab{a}})}]{Slanina:2009JSP}%
  \BibitemOpen
  \bibfield  {author} {\bibinfo {author} {\bibfnamefont {F.}~\bibnamefont
  {Slanina}},\ }\href@noop {} {\bibfield  {journal} {\bibinfo  {journal} {J.
  Stat. Phys.}\ }\textbf {\bibinfo {volume} {135}},\ \bibinfo {pages} {935}
  (\bibinfo {year} {2009}{\natexlab{a}})}\BibitemShut {NoStop}%
\bibitem [{\citenamefont {Slanina}(2009{\natexlab{b}})}]{Slanina:2009PRE}%
  \BibitemOpen
  \bibfield  {author} {\bibinfo {author} {\bibfnamefont {F.}~\bibnamefont
  {Slanina}},\ }\href@noop {} {\bibfield  {journal} {\bibinfo  {journal} {Phys.
  Rev. E}\ }\textbf {\bibinfo {volume} {80}},\ \bibinfo {pages} {061135}
  (\bibinfo {year} {2009}{\natexlab{b}})}\BibitemShut {NoStop}%
\bibitem [{\citenamefont {Chaudhuri}\ and\ \citenamefont
  {Dhar}(2011)}]{Chaudhuri/Dhar:2011}%
  \BibitemOpen
  \bibfield  {author} {\bibinfo {author} {\bibfnamefont {D.}~\bibnamefont
  {Chaudhuri}}\ and\ \bibinfo {author} {\bibfnamefont {A.}~\bibnamefont
  {Dhar}},\ }\href@noop {} {\bibfield  {journal} {\bibinfo  {journal} {EPL}\
  }\textbf {\bibinfo {volume} {94}},\ \bibinfo {pages} {30006} (\bibinfo {year}
  {2011})}\BibitemShut {NoStop}%
\bibitem [{\citenamefont {Chaudhuri}\ \emph {et~al.}(2015)\citenamefont
  {Chaudhuri}, \citenamefont {Raju},\ and\ \citenamefont
  {Dhar}}]{Chaudhuri/etal:2015}%
  \BibitemOpen
  \bibfield  {author} {\bibinfo {author} {\bibfnamefont {D.}~\bibnamefont
  {Chaudhuri}}, \bibinfo {author} {\bibfnamefont {A.}~\bibnamefont {Raju}}, \
  and\ \bibinfo {author} {\bibfnamefont {A.}~\bibnamefont {Dhar}},\ }\href@noop
  {} {\bibfield  {journal} {\bibinfo  {journal} {Phys. Rev. E}\ }\textbf
  {\bibinfo {volume} {91}},\ \bibinfo {pages} {{050103(R)}} (\bibinfo {year}
  {2015})}\BibitemShut {NoStop}%
\bibitem [{\citenamefont {Der\'enyi}\ and\ \citenamefont
  {Vicsek}(1995)}]{Derenyi/Viscek:1995}%
  \BibitemOpen
  \bibfield  {author} {\bibinfo {author} {\bibfnamefont {I.}~\bibnamefont
  {Der\'enyi}}\ and\ \bibinfo {author} {\bibfnamefont {T.}~\bibnamefont
  {Vicsek}},\ }\href@noop {} {\bibfield  {journal} {\bibinfo  {journal} {Phys.
  Rev. Lett.}\ }\textbf {\bibinfo {volume} {75}},\ \bibinfo {pages} {374}
  (\bibinfo {year} {1995})}\BibitemShut {NoStop}%
\bibitem [{\citenamefont {Derenyi}\ and\ \citenamefont
  {Ajdari}(1996)}]{Derenyi/Ajdari:1996}%
  \BibitemOpen
  \bibfield  {author} {\bibinfo {author} {\bibfnamefont {I.}~\bibnamefont
  {Derenyi}}\ and\ \bibinfo {author} {\bibfnamefont {A.}~\bibnamefont
  {Ajdari}},\ }\href@noop {} {\bibfield  {journal} {\bibinfo  {journal} {Phys.
  Rev. E}\ }\textbf {\bibinfo {volume} {54}},\ \bibinfo {pages} {R5} (\bibinfo
  {year} {1996})}\BibitemShut {NoStop}%
\bibitem [{\citenamefont {Rana}\ \emph {et~al.}(2018)\citenamefont {Rana},
  \citenamefont {Goswami}, \citenamefont {Chatterjee},\ and\ \citenamefont
  {Pradhan}}]{Rana/etal:2018}%
  \BibitemOpen
  \bibfield  {author} {\bibinfo {author} {\bibfnamefont {S.}~\bibnamefont
  {Rana}}, \bibinfo {author} {\bibfnamefont {S.}~\bibnamefont {Goswami}},
  \bibinfo {author} {\bibfnamefont {S.}~\bibnamefont {Chatterjee}}, \ and\
  \bibinfo {author} {\bibfnamefont {P.}~\bibnamefont {Pradhan}},\ }\href@noop
  {} {\bibfield  {journal} {\bibinfo  {journal} {Phys. Rev. E}\ }\textbf
  {\bibinfo {volume} {98}},\ \bibinfo {pages} {052142} (\bibinfo {year}
  {2018})}\BibitemShut {NoStop}%
\bibitem [{Note1()}]{Note1}%
  \BibitemOpen
  \bibinfo {note} {To implement the periodic boundary conditions, we assume an
  ordered initial configuration $0 \leq x_{1} \leq x_{2} \protect \ldots \leq
  x_N < L$, and introduce two fictive particles with enslaved coordinates
  $x_0=x_N-L$ and $x_{N+1} = x_1+L$, which implies $x_N - x_1 < L-\sigma $.
  This implementation, where the $x_i$ can assume any real value, is convenient
  for discussing transformation properties of the Langevin equations.
  Alternatively, one could consider the particle positions to be confined to a
  ring of size $L$, corresponding to a mapping $x_i \to x_i \protect \tmspace
  +\thinmuskip {.1667em} \protect \mathrm {mod} \protect \tmspace +\thinmuskip
  {.1667em} L$. However, then the sequential order of the particles can no
  longer be expressed by $x_{1} \leq x_{2} \protect \ldots \leq x_N$ in the
  course of time.}\BibitemShut {Stop}%
\bibitem [{\citenamefont {Tao}\ \emph {et~al.}(2006)\citenamefont {Tao},
  \citenamefont {den Otter}, \citenamefont {Dhont},\ and\ \citenamefont
  {Briels}}]{Tao/etal:2006}%
  \BibitemOpen
  \bibfield  {author} {\bibinfo {author} {\bibfnamefont {Y.-G.}\ \bibnamefont
  {Tao}}, \bibinfo {author} {\bibfnamefont {W.~K.}\ \bibnamefont {den Otter}},
  \bibinfo {author} {\bibfnamefont {J.~K.~G.}\ \bibnamefont {Dhont}}, \ and\
  \bibinfo {author} {\bibfnamefont {W.~J.}\ \bibnamefont {Briels}},\
  }\href@noop {} {\bibfield  {journal} {\bibinfo  {journal} {J. Chem. Phys.}\
  }\textbf {\bibinfo {volume} {124}},\ \bibinfo {pages} {134906} (\bibinfo
  {year} {2006})}\BibitemShut {NoStop}%
\bibitem [{\citenamefont {Scala}\ \emph {et~al.}(2007)\citenamefont {Scala},
  \citenamefont {Voigtmann},\ and\ \citenamefont
  {De~Michele}}]{Scala/etal:2007}%
  \BibitemOpen
  \bibfield  {author} {\bibinfo {author} {\bibfnamefont {A.}~\bibnamefont
  {Scala}}, \bibinfo {author} {\bibfnamefont {T.}~\bibnamefont {Voigtmann}}, \
  and\ \bibinfo {author} {\bibfnamefont {C.}~\bibnamefont {De~Michele}},\
  }\href@noop {} {\bibfield  {journal} {\bibinfo  {journal} {J. Chem. Phys.}\
  }\textbf {\bibinfo {volume} {126}},\ \bibinfo {pages} {134109} (\bibinfo
  {year} {2007})}\BibitemShut {NoStop}%
\bibitem [{\citenamefont {Scala}(2012)}]{Scala:2012}%
  \BibitemOpen
  \bibfield  {author} {\bibinfo {author} {\bibfnamefont {A.}~\bibnamefont
  {Scala}},\ }\href@noop {} {\bibfield  {journal} {\bibinfo  {journal} {Phys.
  Rev. E}\ }\textbf {\bibinfo {volume} {86}},\ \bibinfo {pages} {026709}
  (\bibinfo {year} {2012})}\BibitemShut {NoStop}%
\bibitem [{\citenamefont {Behringer}\ and\ \citenamefont
  {Eichhorn}(2012)}]{Behringer/Eichhorn:2012}%
  \BibitemOpen
  \bibfield  {author} {\bibinfo {author} {\bibfnamefont {H.}~\bibnamefont
  {Behringer}}\ and\ \bibinfo {author} {\bibfnamefont {R.}~\bibnamefont
  {Eichhorn}},\ }\href@noop {} {\bibfield  {journal} {\bibinfo  {journal} {J.
  Chem. Phys.}\ }\textbf {\bibinfo {volume} {137}},\ \bibinfo {pages} {164108}
  (\bibinfo {year} {2012})}\BibitemShut {NoStop}%
\bibitem [{\citenamefont {Seifert}(2012)}]{Seifert:2012}%
  \BibitemOpen
  \bibfield  {author} {\bibinfo {author} {\bibfnamefont {U.}~\bibnamefont
  {Seifert}},\ }\href@noop {} {\bibfield  {journal} {\bibinfo  {journal} {Rep.
  Prog. Phys.}\ }\textbf {\bibinfo {volume} {75}},\ \bibinfo {pages} {126001}
  (\bibinfo {year} {2012})}\BibitemShut {NoStop}%
\bibitem [{\citenamefont {Ambegaokar}\ and\ \citenamefont
  {Halperin}(1969)}]{Ambegoakar/Halperin:1969}%
  \BibitemOpen
  \bibfield  {author} {\bibinfo {author} {\bibfnamefont {V.}~\bibnamefont
  {Ambegaokar}}\ and\ \bibinfo {author} {\bibfnamefont {B.~I.}\ \bibnamefont
  {Halperin}},\ }\href@noop {} {\bibfield  {journal} {\bibinfo  {journal}
  {Phys. Rev. Lett.}\ }\textbf {\bibinfo {volume} {22}},\ \bibinfo {pages}
  {1364} (\bibinfo {year} {1969})}\BibitemShut {NoStop}%
\bibitem [{\citenamefont {Ryabov}\ \emph {et~al.}(2019)\citenamefont {Ryabov},
  \citenamefont {Lips},\ and\ \citenamefont {Maass}}]{Ryabov/etal:2019}%
  \BibitemOpen
  \bibfield  {author} {\bibinfo {author} {\bibfnamefont {A.}~\bibnamefont
  {Ryabov}}, \bibinfo {author} {\bibfnamefont {D.}~\bibnamefont {Lips}}, \ and\
  \bibinfo {author} {\bibfnamefont {P.}~\bibnamefont {Maass}},\ }\href@noop {}
  {\bibfield  {journal} {\bibinfo  {journal} {J. Phys. Chem. C}\ }\textbf
  {\bibinfo {volume} {123}},\ \bibinfo {pages} {5714} (\bibinfo {year}
  {2019})}\BibitemShut {NoStop}%
\bibitem [{\citenamefont {Sch{\"o}nherr}\ and\ \citenamefont
  {Sch{\"u}tz}(2004)}]{Schoenherr/Schuetz:2004}%
  \BibitemOpen
  \bibfield  {author} {\bibinfo {author} {\bibfnamefont {G.}~\bibnamefont
  {Sch{\"o}nherr}}\ and\ \bibinfo {author} {\bibfnamefont {G.~M.}\ \bibnamefont
  {Sch{\"u}tz}},\ }\href@noop {} {\bibfield  {journal} {\bibinfo  {journal} {J.
  Phys. A: Math. Gen.}\ }\textbf {\bibinfo {volume} {37}},\ \bibinfo {pages}
  {8215} (\bibinfo {year} {2004})}\BibitemShut {NoStop}%
\bibitem [{\citenamefont {Percus}(1976)}]{Percus:1976}%
  \BibitemOpen
  \bibfield  {author} {\bibinfo {author} {\bibfnamefont {J.~K.}\ \bibnamefont
  {Percus}},\ }\href@noop {} {\bibfield  {journal} {\bibinfo  {journal} {J.
  Stat. Phys.}\ }\textbf {\bibinfo {volume} {15}},\ \bibinfo {pages} {505}
  (\bibinfo {year} {1976})}\BibitemShut {NoStop}%
\bibitem [{\citenamefont {Marconi}\ and\ \citenamefont
  {Tarazona}(1999)}]{Marconi/Tarazona:1999}%
  \BibitemOpen
  \bibfield  {author} {\bibinfo {author} {\bibfnamefont {U.~M.~B.}\
  \bibnamefont {Marconi}}\ and\ \bibinfo {author} {\bibfnamefont
  {P.}~\bibnamefont {Tarazona}},\ }\href@noop {} {\bibfield  {journal}
  {\bibinfo  {journal} {J. Chem. Phys.}\ }\textbf {\bibinfo {volume} {110}},\
  \bibinfo {pages} {8032} (\bibinfo {year} {1999})}\BibitemShut {NoStop}%
\bibitem [{\citenamefont {Marconi}\ and\ \citenamefont
  {Tarazona}(2000)}]{Marconi/Tarazona:2000}%
  \BibitemOpen
  \bibfield  {author} {\bibinfo {author} {\bibfnamefont {U.~M.~B.}\
  \bibnamefont {Marconi}}\ and\ \bibinfo {author} {\bibfnamefont
  {P.}~\bibnamefont {Tarazona}},\ }\href@noop {} {\bibfield  {journal}
  {\bibinfo  {journal} {J. Phys.: Condens. Mat.}\ }\textbf {\bibinfo {volume}
  {12}},\ \bibinfo {pages} {A413} (\bibinfo {year} {2000})}\BibitemShut
  {NoStop}%
\bibitem [{\citenamefont {Sch{\"u}tz}\ and\ \citenamefont
  {Domany}(1993)}]{Schuetz/Domany:1993}%
  \BibitemOpen
  \bibfield  {author} {\bibinfo {author} {\bibfnamefont {G.}~\bibnamefont
  {Sch{\"u}tz}}\ and\ \bibinfo {author} {\bibfnamefont {E.}~\bibnamefont
  {Domany}},\ }\href@noop {} {\bibfield  {journal} {\bibinfo  {journal} {J.
  Stat. Phys.}\ }\textbf {\bibinfo {volume} {72}},\ \bibinfo {pages} {277}
  (\bibinfo {year} {1993})}\BibitemShut {NoStop}%
\bibitem [{\citenamefont {Maass}\ \emph {et~al.}(2018)\citenamefont {Maass},
  \citenamefont {Dierl},\ and\ \citenamefont {Wolff}}]{Maass/etal:2018}%
  \BibitemOpen
  \bibfield  {author} {\bibinfo {author} {\bibfnamefont {P.}~\bibnamefont
  {Maass}}, \bibinfo {author} {\bibfnamefont {M.}~\bibnamefont {Dierl}}, \ and\
  \bibinfo {author} {\bibfnamefont {M.}~\bibnamefont {Wolff}},\ }\enquote
  {\bibinfo {title} {On {P}hase {T}ransitions in {B}iased {D}iffusion of
  {I}nteracting {P}articles},}\ \ (\bibinfo  {publisher} {Springer
  International Publishing},\ \bibinfo {address} {Cham},\ \bibinfo {year}
  {2018})\ Chap.~\bibinfo {chapter} {9}, pp.\ \bibinfo {pages}
  {147--168}\BibitemShut {NoStop}%
\bibitem [{Note2()}]{Note2}%
  \BibitemOpen
  \bibinfo {note} {In most simulations, the period-averaged density profile
  $\rho _i$ turned out to be monotonically varying with $i$. Only for very high
  $\alpha _{\protect \rm \scriptscriptstyle L}$ and $\alpha _{\protect \rm
  \scriptscriptstyle R}$ small dips appear close to the system boundaries. Then
  $\rho _{\protect \rm \scriptscriptstyle L}$ and $\rho _{\protect \rm
  \scriptscriptstyle R}$ were determined from the region of monotonically
  varying profile.}\BibitemShut {Stop}%
\bibitem [{\citenamefont {Kolomeisky}(1998)}]{Kolomeisky:1998}%
  \BibitemOpen
  \bibfield  {author} {\bibinfo {author} {\bibfnamefont {A.~B.}\ \bibnamefont
  {Kolomeisky}},\ }\href@noop {} {\bibfield  {journal} {\bibinfo  {journal} {J.
  Phys. A: Math. Gen.}\ }\textbf {\bibinfo {volume} {31}},\ \bibinfo {pages}
  {1153} (\bibinfo {year} {1998})}\BibitemShut {NoStop}%
\bibitem [{\citenamefont {Brankov}\ \emph {et~al.}(2004)\citenamefont
  {Brankov}, \citenamefont {Pesheva},\ and\ \citenamefont
  {Bunzarova}}]{Brankov/etal:2004}%
  \BibitemOpen
  \bibfield  {author} {\bibinfo {author} {\bibfnamefont {J.}~\bibnamefont
  {Brankov}}, \bibinfo {author} {\bibfnamefont {N.}~\bibnamefont {Pesheva}}, \
  and\ \bibinfo {author} {\bibfnamefont {N.}~\bibnamefont {Bunzarova}},\
  }\href@noop {} {\bibfield  {journal} {\bibinfo  {journal} {Phys. Rev. E}\
  }\textbf {\bibinfo {volume} {69}},\ \bibinfo {pages} {066128} (\bibinfo
  {year} {2004})}\BibitemShut {NoStop}%
\end{thebibliography}

%

\end{document}